\documentclass[useAMS,usenatbib,twocolumn]{mn2e}
\usepackage{xspace}
\usepackage{hyperref}
\usepackage{aas_macros}
\usepackage{url}
\usepackage{amsmath}
\usepackage{color}
\usepackage{amssymb}
\usepackage{multirow}
\usepackage[capitalise]{cleveref}
\usepackage{graphicx}

\newcommand{\VLA}{VLA\xspace}

\newcommand{\clean}{\textsc{CLEAN}\xspace}
\newcommand{\pybdsf}{\textsc{PyBDSF}\xspace}
\newcommand{\casa}{\textsc{CASA}\xspace}
\newcommand{\sersic}{S\'{e}rsic\xspace}
\newcommand{\imshape}{{\sc{IM3SHAPE}}\xspace}

\newcommand{\emerlin}{\emph{e}-MERLIN\xspace}

\newcommand{\eint}{\epsilon^{\mathrm{int}}}

\newcommand{\degsq}{\, \mathrm{deg}^2}

\definecolor{mygrey}{gray}{0.6}
\newcommand{\ih}[1]{{\textcolor{black}{#1}}}

\newcommand{\jbca}{{Jodrell Bank Centre for Astrophysics, Department of Physics \& Astronomy, The University of Manchester, Manchester M13 9PL, UK}}

\title[SuperCLASS - III.  Radio \& optical weak lensing]{SuperCLASS -- III. Weak lensing from radio and optical observations in Data Release 1}
\author[Harrison et al.]{Ian Harrison\textsuperscript{\thanks{E-mail: ian.harrison-2@manchester.ac.uk}}$^{1,2}$,
Michael L. Brown$^{1}$,
Ben Tunbridge$^{1}$,
Daniel B. Thomas$^{1}$, \newauthor
Tom Hillier$^{1}$,
A. P, Thomson$^{1}$,
Lee Whittaker$^{1,3}$,
Filipe B. Abdalla$^{3}$, \newauthor
Richard A. Battye$^{1}$,
Anna Bonaldi$^{1,4}$,
Stefano Camera$^{1,5,6}$, 
Caitlin M. Casey$^{7}$, \newauthor
Constantinos Demetroullas$^{1,8}$,
Christopher A. Hales$^{9,10}$, 
Neal J. Jackson$^{1}$, \newauthor
Scott T. Kay$^{1}$, 
Sinclaire M. Manning$^{7}$, 
Aaron Peters$^{1}$,
Christopher J. Riseley$^{11,12,13}$, \newauthor
Robert A. Watson$^{1}$, (SuperCLASS Collaboration)\\
$^{1}$ \jbca\\
$^{2}$ Department of Physics, University of Oxford, Denys Wilkinson Building, Keble Road, Oxford OX1 3RH, UK\\
$^{3}$ Department of Physics and Astronomy, University College London, Gower Street, London, WC1E 6BT, UK\\
$^{4}$ SKA Organization, Jodrell Bank, Lower Whitington, Macclesfield, SK11 9DL, UK\\
$^5$ Dipartimento di Fisica, Universit\'a degli Studi di Torino, Via P. Giuria 1, 10125 Torino, Italy\\
$^6$ INFN -- Istituto Nazionale di Fisica Nucleare, Sezione di Torino, Via P. Giuria 1, 10125 Torino, Italy\\
$^7$ Department of Astronomy, University of Texas at Austin, 2515 Speedway Blvd, Stop C1400, Austin, Texas, U.S.A. \\
$^8$ Cyprus University of Technology, Archiepiskopou Kyprianou 30, Limassol, 3036, Cyprus\\
$^9$ National Radio Astronomy Observatory, PO Box 0, Socorro, NM 87801, USA\\
$^{10}$ School of Mathematics, Statistics and Physics, Newcastle University, Newcastle upon Tyne NE1 7RU, UK\\
$^{11}$ Dipartimento di Fisica e Astronomia, Universit\`a degli Studi di Bologna, via P. Gobetti 93/2, 40129 Bologna, Italy\\
$^{12}$ INAF -- Istituto di Radioastronomia, via P. Gobetti 101, 40129 Bologna, Italy\\
$^{13}$ CSIRO Astronomy and Space Science, PO Box 1130, Bentley, WA 6102, Australia
}
\begin{document}

\pagerange{\pageref{firstpage}--\pageref{lastpage}} \pubyear{2020}

\maketitle

\label{firstpage}

\begin{abstract}
We describe the first results on weak gravitational lensing from the SuperCLASS survey: the first survey specifically designed to measure the weak lensing effect in radio-wavelength data, both alone and in cross-correlation with optical data. We analyse $1.53 \degsq$ of optical data from the Subaru telescope and $0.26 \degsq$ of radio data from the \emerlin and VLA telescopes (the DR1 data set). Using standard methodologies on the optical data only we make a significant (\ih{$10\sigma$}) detection of the weak lensing signal (a shear power spectrum) due to the massive \ih{supercluster} of galaxies in the targeted region. For the radio data we develop a new method to measure the shapes of galaxies from the interferometric data, and we construct a simulation pipeline to validate this method. We then apply this analysis to our radio observations, treating the \emerlin and VLA data independently. We achieve source densities of $0.5 \,$arcmin$^{-2}$ in the VLA data and $0.06 \,$arcmin$^{-2}$ in the \emerlin data, numbers which prove too small to allow a detection of a weak lensing signal in either the radio data alone or in cross-correlation with the optical data. Finally, we show preliminary results from a visibility-plane combination of the data from \emerlin and VLA which will be used for the forthcoming full SuperCLASS data release. This approach to data combination is expected to enhance both the number density of weak lensing sources available, and the fidelity with which their shapes can be measured.
\end{abstract}

\section{Introduction}
\label{sec:introduction}
Observations of weak gravitational lensing in the optical wavebands have in the past twenty years moved from first detections of the signal \citep{2000MNRAS.318..625B,2000astro.ph..3338K,2000A&A...358...30V,2000Natur.405..143W} to competitive cosmological constraints (KiDS-450, \citealt{2017MNRAS.465.1454H}; DES-Y1, \citealt{2018PhRvD..98d3528T}; HSC first-year, \citealt{2018arXiv180909148H}). The current generation of experiments has begun to show a consistent picture of cosmological structure formation at late times, and their final results will provide useful information on the apparent tensions between such probes and primary CMB measurements \citep[see e.g.][for an overview]{2018arXiv180604649R}. Following this, the next generation of weak lensing experiments such as LSST \citep{2018arXiv180901669T} and the \emph{Euclid} satellite \citep{2018LRR....21....2A}, will further reduce statistical errors by an order of magnitude. However, there is concern that systematic errors may already be overwhelming the statistical error bars. Cosmic shear surveys rely on compilations of shapes, fluxes and distances for between millions and billions of galaxies, meaning even small biases in analysis pipelines may compound to have significant impact on inferred cosmological parameters. One potential way to mitigate these problems is through performing weak lensing observations in the radio waveband. As opposed to the filled apertures and CCDs of optical and near-IR telescopes, the relevant radio telescopes are interferometers. By using the small and intermediate spatial scales available from interferometer data to measure galaxy shapes, we expect systematic errors on the measurement of the weak lensing signal that are caused by the instrument to be uncorrelated between the radio and the optical. By cross-correlating maps of the weak lensing signal between the two wavebands, systematics can be removed \citep{2017MNRAS.464.4747C} but statistical constraining power is conserved \citep{2016MNRAS.463.3674H}. Surveys possible with the first phase Square Kilometre Array \citep[SKA1;][]{2018arXiv181102743S}, which is expected to begin observing near the end of the next decade, will be comparable in weak lensing constraining power to the current optical surveys \citep{2016MNRAS.463.3686B}, meaning their cross-correlation combinations will be both statistically precise and systematically robust.

However, radio surveys have thus far lacked the combination of sub-arcsecond resolution, sub-$\mu$Jy depth and wide ($>10 \degsq$) area necessary to detect high number densities ($>1 \,$arcmin$^{-2}$) of star-forming galaxies at $z\sim1$ and hence make a firm detection of the weak lensing signal. These two tentative detections to date have both used observations not designed for weak lensing science. \cite{2004ApJ...617..794C} used the wide ($\sim10^4\, \degsq$) but shallow ($\sim 5 \times 10^{-3}\,$gal arcmin$^{-2}$) FIRST survey at $1.4\,$GHz to make a $3.6\sigma$ detection of aperture mass variance on angular scales of $1$--$4$ degrees, whilst \cite{2016MNRAS.456.3100D} combined the FIRST radio and SDSS optical surveys to measure a shear signal from cross-correlations of radio and optical shapes at $2.7\sigma$. By correlating the shapes of FIRST radio galaxies with the positions of SDSS DR10 objects \cite{2018MNRAS.473..937D} were also able to make a firm ($10\sigma$) detection of a galaxy-galaxy lensing signal. Studies of the COSMOS field at both $1.4\,$GHz \citep{2016MNRAS.463.3339T} and $3\,$GHz \citep{2019MNRAS.488.5420H} radio frequencies have also attempted to detect the optical cross-correlation signal, but have not achieved sufficient number densities to make firm detections.

The subject of this paper is the SuperCLASS (Super CLuster Assisted Shear Survey) survey \citep[][hereafter Paper I]{battye2019}, which is a legacy survey using the \emerlin telescope, and is the first survey designed from the outset for making a detection of the weak lensing signal with radio observations. In the companion paper (Paper I) we describe the full multiwavelength data set taken for SuperCLASS, along with the details of the data reduction and basic science results for the Data Release 1 (DR1) subset. \cite{manning2019} (hereafter Paper II) describes the use of these observations to constrain the photometric redshift distribution of the optical sources and understand the spectral energy distributions of matched radio sources. In this paper we use these observations to attempt to measure a weak lensing signal. Limiting ourselves to the $0.26\,\degsq$ DR1 region of the SuperCLASS field, we demonstrate what can be achieved using measurements of star-forming galaxy shapes from the VLA and \emerlin radio data and Subaru optical data. With these measurements we place constraints on the weak lensing signal as quantified by the radio, optical and radio-optical-cross power spectra. We validate our shape measurement methods for radio galaxies on simulations of the data set with known inputs, and show that the data which will be available for the full survey of $\sim1\,\degsq$ may be capable of a detection of a radio-optical weak lensing cross-correlation signal.

For a comprehensive introduction to and review of weak lensing cosmology we refer the reader to \cite{2015RPPh...78h6901K}.

In \cref{sec:survey} we briefly introduce the survey and the radio and optical observations. \Cref{sec:optical_meas} then details the creation of the optical shape catalogue, and \cref{sec:radio_meas} the creation of the radio shape catalogue from our VLA and \emerlin data, whose properties we describe in \cref{sec:radio_meas-shapes}. We then present our measurement of the radio, optical and radio-optical-cross shear power spectra in the SuperCLASS DR1 region in \cref{sec:power_spectra}. We present the method of data combination between VLA and \emerlin data sets to be used for the full SuperCLASS survey in \cref{sec:data_combination}, and conclude in \cref{sec:summary}.

\section{The SuperCLASS survey}
\label{sec:survey}
For a full description of the observations making up the SuperCLASS Data Release 1 (DR1) data set, we refer the reader to \ih{Paper I}. We will briefly describe the key points here. The full SuperCLASS field, displayed in \cref{fig:superclass_field}, consists of a $\sim 1.53 \degsq$ region of the Northern sky around 10h 15m RA, +68d Declination. This field contains five candidate Abell galaxy clusters of mass $\sim 10^{14}\,M_{\odot}$ at redshift $z=0.2$ -- the supercluster referenced in the SuperCLASS acronym -- which we expect to enhance the lensing signal available to be measured in the region by a factor of $\sim 2$ \citep{2018MNRAS.474.3173P}. This region has been covered in a number of different observations for the SuperCLASS project (see Paper I), but here we focus on the three data sets used for the weak lensing science analysis: \emerlin and VLA radio data, and Subaru Suprime-Cam (SC) and Hyper-Suprime-Cam (HSC) optical data. For our primary analysis (which relies on the available radio data) we also restrict ourselves to a $0.26\,\degsq$ region indicated in \cref{fig:superclass_field} and referred to as the DR1 region. Though observations are complete for the full SuperCLASS data set, and optical data are fully reduced, the DR1 region is the one which currently has radio data which is reduced to a uniform depth and ready for science analysis.
\subsection{Radio observations}
\label{sec:survey-radio}
\subsubsection{VLA observations}
\label{sec:survey-radio-VLA}
The full SuperCLASS field was observed in 24 hours of A-configuration Karl G. Jansky Very Large Array (VLA or JVLA) time in August 2015. \ih{A total of 112 separate telescope pointings were taken. These pointings were distributed in an interlocking hexagonal strategy with $5.7^{\prime}$ between each centre. This enabled the creation of a mosaic image with approximately uniform RMS noise.} The frequency range was 1--2$\,$GHz (L-band), divided in 4,000 spectral channels of 250$\,$kHz each, with a 1 second time sampling. Each pointing was individually imaged using the \casa \citep{2007ASPC..376..127M} {\tt tclean} task, with a $1.4\,\degsq$ field of view of $0.2\,$arcsec$^2$ pixels. Visibility plane data had Radio Frequency Interference (RFI) removed and calibrations applied. The restoring beam was fixed at $1.9 \times 1.5$ arcsec, with a position angle $80\,\deg$ East from North. During imaging deconvolution the Briggs weighting scheme was applied to the visibility data to balance the desire for a low image-plane noise level and low levels of PSF sidelobes, and wide field corrections were applied. The {\tt tclean} algorithm was also applied in the multi-scale cleaning mode, using scales of 0, 4 and 12 arcseconds. Individual pointings were then mosaiced together using a slant orthographic SIN projection, reducing the image-plane noise level from 20 to 7$\, \mu$Jy$\,$beam$^{-1}$, except for regions around bright contaminating sources, which were not considered for shape measurement. In this work, we make use of only the $0.26\,\degsq$ DR1 region.
\subsubsection{\emerlin observations}
\label{sec:survey-radio-emerlin}
Observations covering the DR1 region were also taken by the \emerlin telescope, over the period 2014-2016. 49 pointings were taken in total, again in a seven point mosaic pattern, with L-band frequency coverage from 1.204 to 1.717$\,$GHz, split into a total of 4,096 frequency channels of 125$\,$kHz each and 1 second time sampling. As with the VLA data, visibility plane data had RFI removed and calibrations applied. These pointings were then imaged using the WSCLEAN \citep{2014MNRAS.444..606O} package, accounting for $w$-projection terms in wide field imaging and using natural weighting of the visibility data to maximise sensitivity. These pointings were again mosaiced together to provide a final image with a uniform noise region in the central $0.26\,\degsq$ with noise RMS $\sim7\, \mu$Jy$\,$beam$^{-1}$. A further 49 pointings have subsequently been observed in the period 2017-2018, covering the rest of the Northern $\sim1\,\degsq$ region of the SuperCLASS field. However, these data have not yet been reduced to a science-ready state and we do not consider them here, deferring their analysis to the next data release.
\subsection{Optical observations}
\label{sec:survey-optical}
Optical observations used here were taken using the Subaru $8.4 \,$m telescope, using the Suprime-Cam (SC) for $BVr^\prime i^\prime$ bands. Six pointings were taken to cover the field, with an average seeing of 1.0 to 1.4 arcsec and time divided across filters to achieve a uniform depth of 25 magnitude in all bands. Data reduction was performed with the Subaru Suprime-Cam Data Reduction and Optical Imaging software \citep[SDFRED2,][]{2004ApJ...611..660O}. Observations were also taken for SuperCLASS in the $z^\prime$ band but these are not included due to the uneven coverage, with only four (of six) fields being observed in total. Photometric redshifts for these sources, in particular the star-forming galaxies which also appear in the \emerlin image, are described in Paper II.
\begin{figure}
 \centering
 \includegraphics[width=0.5\textwidth,angle=0]{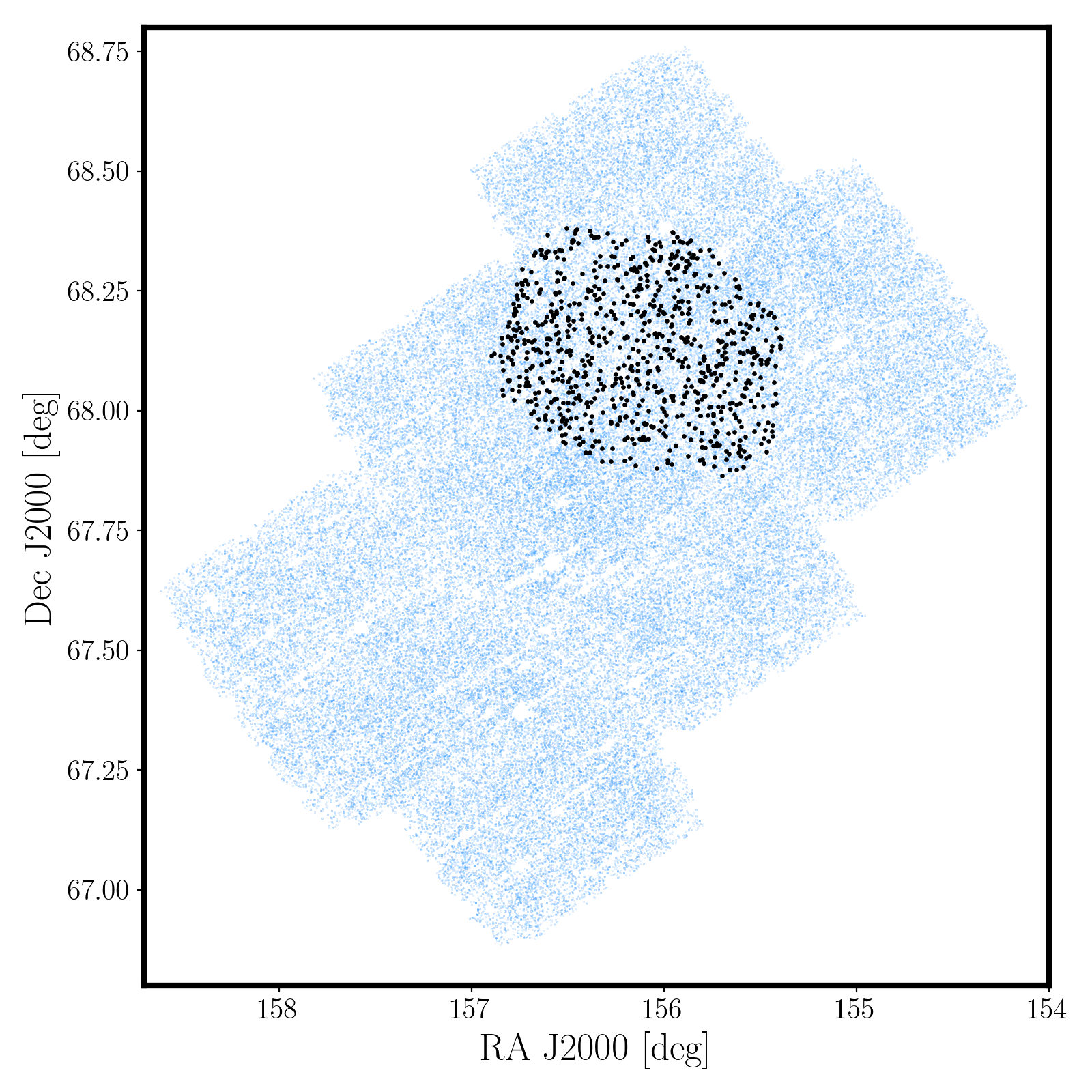}
 \caption[Weak lensing sources]{Map of the weak lensing sources used in this work. The DR1 region is indicated by the radio sources (black), whilst the full SuperCLASS field is shown by the extent of the optical sources (blue).}
 \label{fig:superclass_field}
\end{figure}
\section{Optical shape measurement}
\label{sec:optical_meas}
From our optical observations we wish to select high redshift ($z \gtrsim 0.5$) star-forming galaxies and measure their shapes, a process which we describe here.
\subsection{Weak lensing catalogue}
\label{sec:optical_meas-catalogue}
For source detection, it is desirable to combine the raw observations in such a way as to optimise sensitivity. A co-added image was created consisting of all available and complete bands, $BVr^\prime i^\prime$. The astrometric and photometric calibration was performed using the \textsc{Astrometry.net} software package \citep{2010AJ....139.1782L}. Data from the Second-Generation Guide Star Catalog \citep{2008AJ....136..735L} and the Data Release 1 (DR1) of the Panoramic Survey Telescope And Rapid Response System (PanSTARRS) survey \citep{2016arXiv161205560C} were used to perform this calibration, due to the similar band coverage. The measured magnitudes were then colour-corrected using stellar templates from the Stellar Flux Library \citep{1998PASP..110..863P}, convolved with the Subaru filters.

The co-added image for source finding is PSF smoothed to the poorest seeing conditions of exposures included, which was found to be $1.38''$. This smoothing of the PSF is not ideal for shape measurement since we are limiting the resolving power of all exposures to match the worst seeing. This co-added image was therefore used purely for source finding. A source photometric catalogue was compiled from the co-added image using the source extraction software, \textsc{SExtractor} \citep{1996A&AS..117..393B}.

The full photometric source catalogue from Paper II consists of $\sim 6\times 10^{5}$ sources, but not all of these will be suitable for shape measurement. We perform a number of cuts on various source features to select galaxies on which it is suitable to apply our shape measurement method, which will then be used for weak lensing shear estimation. We cut on: source size to ensure morphological information is available (that the source is resolved); positional offset (of the model fit centroid from the detection centroid) and measured signal-to-noise-ratio (SNR) to ensure we are not fitting spurious detections of noise; and a number of model goodness-of-fit statistics (as described in \cref{tab:selectioncriteria}), which removed failures in fitting due to overly-complex source morphology, partially masked sources, or other failures. These selection criteria are defined in \Cref{tab:selectioncriteria}.

\ih{Cluster members are also removed by cutting sources which have $z_{\rm phot}=0.2\pm0.08$ and $i^\prime<22.5$ (see Paper II, section 4.3.2) from the shape catalogue. The final shape catalogue for further analysis consists of $111,020\:$sources, which corresponds to an optical source density of $n_{\rm gal}^{\rm O} = 19.8\:\mathrm{arcmin}^{-2}$.}
\begin{table}
\caption[Shape Analysis Selection Criteria]{The source selection criteria used to optimise the optical shape catalogue for analysis. We make various cuts on source sizes, position offset (offset in right ascension $\Delta\alpha$ and in declination $\Delta\delta$) detection significance and fit statistics.}
\label{tab:selectioncriteria}
\begin{center}
\begin{tabular}{lc}
\hline
Type & Selection Criteria \\
\hline
 \multirow{2}{*}{Source size} & $\mathrm{FWHM}_{\rm source}\: >\: 1.2 \times \mathrm{FWHM}_{\rm PSF}$ \\
  & $\mathrm{FWHM}_{\rm source}\: <\: 6\ \mathrm{arcsec}$ \\ [0.4cm]
 \multirow{2}{*}{Position} & $\Delta\alpha\:<\:1.0\: \mathrm{arcsec}$ \\
  &  $\Delta\delta\:<\:1.0\:\mathrm{arcsec}$ \\ [0.4cm]
  Detection & $20\: < \: \mathrm{SNR} \: < \: 10^{5}$ \\ [0.4cm]
  \multirow{4}{*}{Goodness-of-fit statistics} &  $\mathrm{ellipticity\ modulus}\: <\: 0.95$ \\
  & $\mathrm{model\ minimum}\: >\: -0.05$ \\
  & $\mathrm{iterations}\: < 500$ \\
  & $ 0.5\:<\:\chi^{2}_{\mathrm{pixel}}\:<\:2.5$ \\
\hline
\end{tabular}
\end{center}
\end{table}
\subsection{PSF estimation}
\label{sec:optical_meas-psf}
In order that we may deconvolve their effect on source shapes, PSF models are constructed from the population of stars visible across the field. Unlike galaxies, stars are intrinsically point-like from the point of view of these observations and thus provide the approximate PSF response on an irregular grid of positions on the sky. By constructing a stellar catalogue for each exposure, we can build the required PSF models. In order to extract the best PSF models possible, we perform PSF estimation (and subsequent galaxy shape measurement) using the individual exposure images, rather than the co-added and smoothed images used for source detection in \cref{sec:optical_meas-catalogue} above.

\begin{figure}
 \centering
 \includegraphics[width=0.5\textwidth,angle=0]{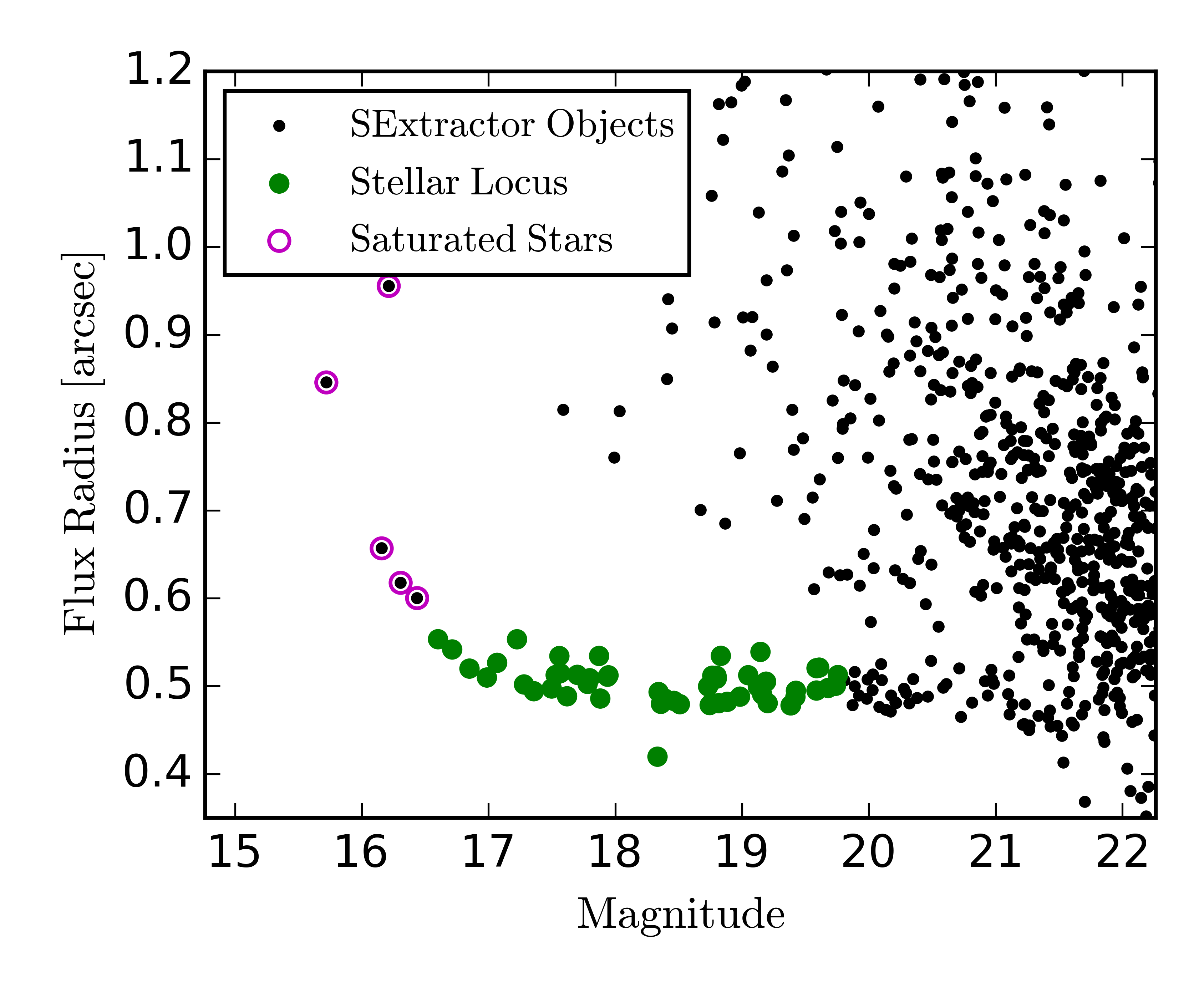}
 \caption[Optical Stellar Locus]{Example of stellar locus identification. The green points show the successfully identified members of the stellar locus. Saturated stars, which are removed from the analysis, are also identified.}
 \label{fig:Optical Stellar Locus}
\end{figure}

The full iterative procedure for identifying the stellar locus for PSF modelling in the source catalogue is outlined below:
\begin{enumerate}
\item \label{n1} \textbf{Calibrated single exposures}. The single exposures were prepared and calibrated from the raw data. We perform astrometric calibration of each CCD exposure independently via a series of steps. The first involves creating an initial source catalogue per exposure using the source extraction software, \textsc{SExtractor} mentioned previously in \cref{sec:optical_meas-catalogue}. Astrometric solutions were then calculated by matching this catalogue to the Second-Generation Guide Star Catalog \citep{2008AJ....136..735L} as our astrometric reference catalogue; the solutions are computed using the observational calibration software \textsc{SCAMP} (Software for Calibrating AstroMetry and Photometry). Finally the astrometric solutions are applied to the single exposure images using the image resampling software package \textsc{SWarp} \citep{2002ASPC..281..228B}.

\item \label{n2} \textbf{Initial stellar catalogue}. An initial source catalogue is constructed using \textsc{SExtractor} on each exposure. We extract the stellar population in this first iteration by applying a selection cut to the \textsc{SExtractor} Neural-Network-based star/galaxy classifier, \texttt{class\_star}, which is described in more detail in \cite{1996A&AS..117..393B}. An initial star selection cut was made with, \texttt{class\_star}$\,>0.9$ and signal-to-noise-ratio SNR$\,> 20$ to produce the first iteration stellar catalogue. The reliability of \texttt{class\_star} is discussed in \cite{2005astro.ph.12139H}.

\item \label{n3} \textbf{Improved stellar catalogue}. The full surface brightness PSF models were calculated using the software package \textsc{PSFex} \citep[PSF Extractor,][]{2011ASPC..442..435B} from the initial stellar catalogue obtained in step (ii). These models were then fed back into a second run of \textsc{SExtractor} to construct a second iteration stellar catalogue. This time the software incorporates the PSF model into the morphology estimates and we are able to produce a more reliable star/galaxy classification. We adapt the pseudo-code detailed in \cite{2016MNRAS.460.2245J} (section 2.2) for the star source selection which we implement as follows:\\

\begin{minipage}{0.7\linewidth}
\begin{small}
\begin{tabular}{p{1.65cm}p{0.2cm}p{5cm}}
\texttt{size\_test} &$=$& $0.9 <$\texttt{flux\_radius}/\texttt{seeing} $< 1.3$\\
 &&and \texttt{mag\_auto} $< 24.0$\\
 &&\\
\texttt{star\_test} &$=$& \texttt{class\_star} $> 0.3$\\
 &&\\
\texttt{locus\_test} &$=$& \texttt{spread\_model}\\
 &&$+$\texttt{spreaderr\_model} $< 0.003$\\
 &&\\
\texttt{faint\_psf\_test} &$=$& \texttt{mag\_psf} $> 40.0$\\
 &&and \texttt{mag\_auto} $< 26.0$\\
 &&and \texttt{mag\_psf} $< 90.0$\\
 &&\\ 
\texttt{stars} &$=$& \texttt{size\_test} and\\
 &&[\texttt{locus\_test} or \texttt{star\_test}]\\
 &&and \texttt{not} \texttt{faint\_psf\_test}\\
\end{tabular}
\end{small}
\end{minipage}\\
\newline
where the $\texttt{seeing}$ is estimated in the initial data reduction. The \texttt{locus\_test} procedure identifies the source locations relative to the stellar locus via the \texttt{spread\_model} parameter. This is an additional star/galaxy classifier based on the difference in goodness-of-fit between the best fitting PSF model and a model made from the same PSF convolved with a circular exponential disc model \citep[see][for a more detailed discussion on the \texttt{spread\_model} classifier]{2012ApJ...757...83D}. Again we only include stars with SNR$\,>20$ in our final stellar sample. A star is included in the final stellar catalogue if it has measured \texttt{SExtractor} properties passing the \texttt{size\_test} (an initial stellar locus approximation); is not flagged as junk by the \texttt{faint\_psf\_test}, and meets either of the \texttt{locus\_test} or \texttt{star\_test}. \cref{fig:Optical Stellar Locus} provides an example of this classification.

\item \label{n4} \textbf{Final PSF models}. Finally we again generate PSF models, this time from the improved stellar sample from step (iii) and using \textsc{PSFex}. This provides a PSF model as a function of position on the sky for each of the individual exposures.
\end{enumerate}

\subsection{PSF model diagnostics}
\label{sec:optical_meas-psf_diagnostics}
To assess the quality of our PSF models constructed in \cref{sec:optical_meas-psf}, we performed a number of systematic checks. In particular we want to ensure that we have minimised any potential PSF contamination of the galaxy shapes and any remaining systematic is sufficiently smaller than our expected signal. These systematic errors may be introduced through inaccuracies in the PSF modelling process, for example through an inappropriate stellar sample or errors in the interpolation. Such errors in the PSF models will be correlated among galaxies and hence become a source of systematic bias in estimated shapes which will propagate into the shear estimates.

A useful diagnostic for quantifying the PSF model error was first introduced by \cite{2010MNRAS.404..350R} and provides a test using the observed shapes of individual stars reproduced by the PSF model constructed at that same location by the model interpolation. This diagnostic is defined as
\begin{equation}
\rho_{1}(\theta) = \langle \delta e^{*}_{\mathrm{PSF}}(\mathrm{x})\: \delta e_{\mathrm{PSF}}(\mathrm{x}+\theta) \rangle,
\end{equation}
where $e^{*}_{\mathrm{PSF}}$ is the PSF model ellipticity \citep[defined as in][equation 25]{2015RPPh...78h6901K} at a given location and $\delta e_{\mathrm{PSF}}$ is the residual between the true measured stellar ellipticity and the PSF model ellipticity evaluated at that location. Note $\langle a^{*} \: a \rangle$ defines the auto-correlation function of a given parameter $a$ \citep[for a full description and set of definitions for angular correlation functions see][section 3.8]{2015RPPh...78h6901K}.

Following the approach taken in \cite{2008A&A...484...67P} and \cite{2015MNRAS.454.3500K}, we also introduce a second statistic relating to the PSF residual sizes:
\begin{equation}
\rho_{2}(\theta) = \langle  \delta R^{2*}_{\mathrm{PSF}}(\mathrm{x})\: \delta R^{2}_{\mathrm{PSF}}(\mathrm{x}+\theta) \rangle,
\end{equation}
where $R^{2}_{\mathrm{PSF}} = Q_{11} + Q_{22}$ are the unweighted second order moments of the PSF model residual (residual from the PSF model and observed star size). 

For the purposes of this work (detection of the angular power spectrum, rather than precision cosmological measurements), anisotropy systematics from PSF model interpolation should be subdominant to the expected shear lensing signal. To quantify this we measure the statistics, $\rho_{1}$ and $\rho_{2}$, from the data. To obtain $\delta e_{\mathrm{PSF}}$ and $\delta R^{2}_{\mathrm{PSF}}$, an estimate of the star population and PSF model shapes and sizes are required. These are obtained from the quadrupole moment measured using the \texttt{FindAdaptiveMom} tool in the GalSim software package \citep{2015A&C....10..121R}. This tool iteratively computes a best fitting elliptical Gaussian model to find the equivalent weighted fit to the image quadrupole moments \citep[see][]{2003MNRAS.343..459H}. The star population shapes and associated PSF shapes at the star positions are obtained directly from the \texttt{FindAdaptiveMom} task applied to the data image and the synthetic PSF model cut-outs, respectively.

\begin{figure}
\centering
  \includegraphics[width=0.5\textwidth]{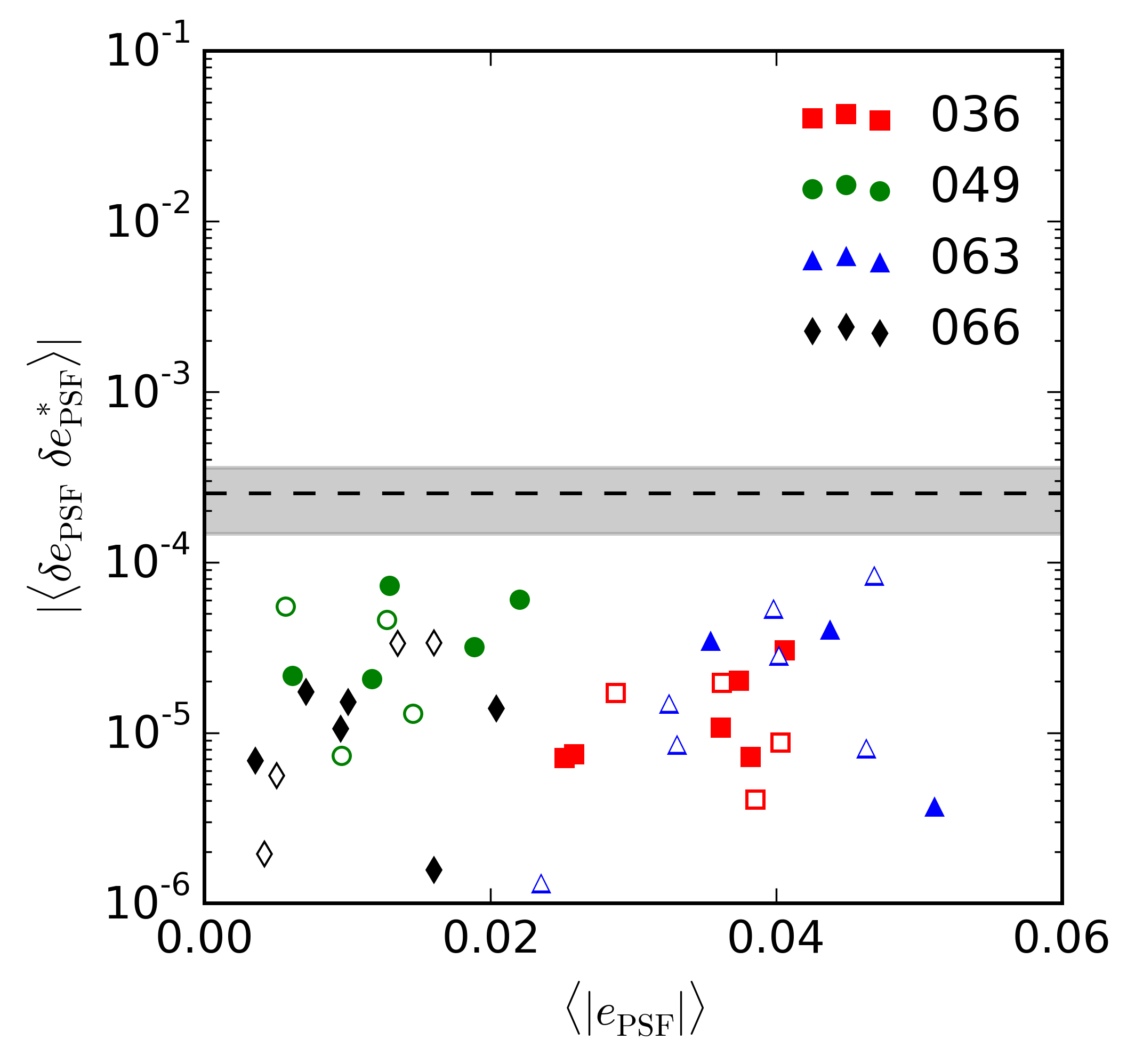}\\
  \includegraphics[width=0.4755\textwidth]{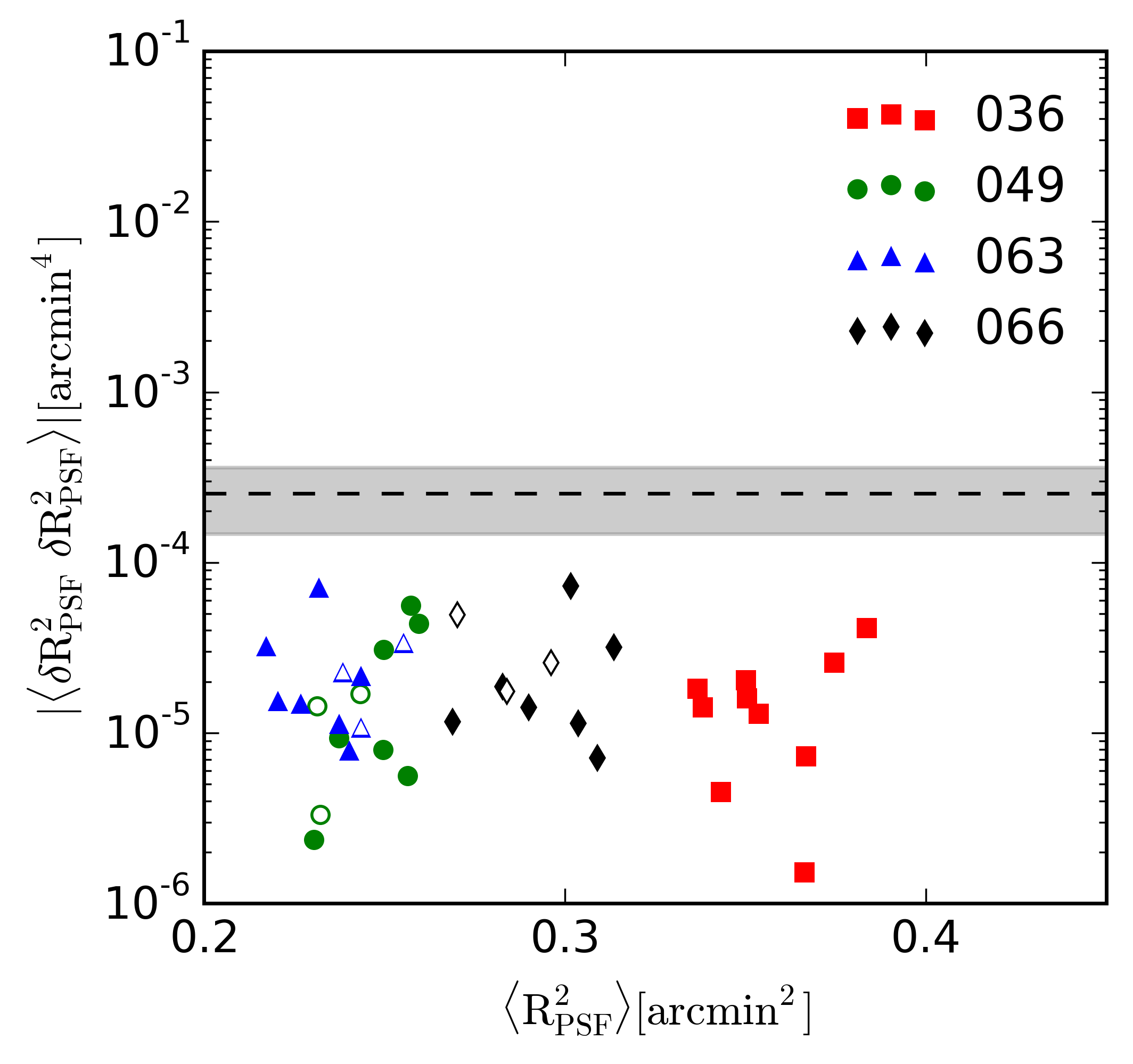}
\caption[$\rho_{1}$ and $\rho_{2}$ PSF Model Interpolation Dianostic]{The PSF model interpolation diagnostics are presented for a selection of SuperCLASS Subaru SC photometric exposures in the observation band $i^\prime$, from the SCLASS2 sub-field as an example. The \textit{top panel} shows the $\delta e_{\rm PSF}$ auto correlation and the \textit{bottom panel} shows $\delta R^{2}_{\rm PSF}$ auto correlation, which relate to the PSF shape and size reproducible from the input stellar population. The different symbols represent $4$ different epochs of observations. Each exposure has $10$ data points, one for each of the CCDs. The dashed line shows the average expected signal from the SuperCLASS-like \ih{supercluster} simulations with the spread across the different clusters indicated by shaded region. The unfilled markers indicate negatives. This shows that PSF residual systematics are constrained to be at a level below 10\% of the expected signal strength.}
\label{fig:deRautocorrelation}
\end{figure}

In \Cref{fig:deRautocorrelation} we show the $\rho_{1}$ and $\rho_{2}$ statistics as a function of the associated parameter from the optical data at a separation of $1\:$arcmin for a selection of exposures in the SCLASS2 sub-field (one of six). A separation of $1\:$arcmin was chosen as in \cite{2015MNRAS.454.3500K}, since this is roughly the limiting scale at which the statistics $\rho_{1}$ and $\rho_{2}$ can be reliably computed, given the stellar source density. We compare $\rho_{1}$ and $\rho_{2}$ with the expected SuperCLASS shear signal (obtained from the \ih{supercluster} $N$-body simulations discussed in \cref{sec:radio_meas-simuclass-shear}). The $\rho_{1}$ and $\rho_{2}$ can be seen to be lower than the expected signal. For the purpose of this analysis, this is sufficient for detection. We also consider an alternative approach to setting requirements on the PSF models, in terms of the PSF leakage (denoted $\alpha$) into galaxy shapes. This provides a post process systematic check, looking for traces of PSF contamination in the measured galaxy shapes.

We adopt the parametrisation of the galaxy shape ellipticity contributions first proposed by \cite{2006MNRAS.368.1323H},
\begin{equation}\label{eq:bias_form}
\langle e \rangle = (1+m)\: \gamma\: +\: \alpha\: e_{\mathrm{PSF}}\: +\: c ,
\end{equation}
where $m$ is the multiplicative error, $c$ is the additive error and $\alpha$ defines the PSF leakage upon the galaxy shapes. The PSF leakage term $\alpha$ is often combined with the additive bias $c$. However we follow \cite{2016MNRAS.460.2245J} and keep $\alpha$ decoupled from $c$ since we are interested in testing the PSF leakage explicitly. The minimisation of these three quantities is the objective of the systematic corrections applied to weak lensing data sets.

By substituting \cref{eq:bias_form} into the shear two-point correlation functions, an expression for the systematic error on the two correlation functions is obtained:
\begin{equation}\label{eq:syst}
\delta\xi_{\pm}(\theta) \simeq 2 m \xi_{\pm}(\theta) + \alpha^{2} \xi^{pp}_{\pm} (\theta) + \xi^{cc}_{\pm} (\theta),
\end{equation}
where $\xi^{pp}_{\pm}$ is the auto-correlation function of the PSF shapes, and $\xi^{cc}_{\pm}$ is the auto-correlation function of the additive error, $c$. Here, we have assumed that the systematic quantities ($m$, $c$ and $\alpha$) are uncorrelated.

\begin{figure}
\centering
  \includegraphics[width=0.5\textwidth]{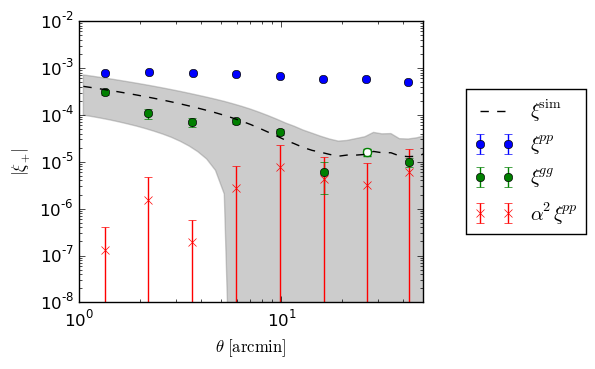}\\
  \includegraphics[width=0.5\textwidth]{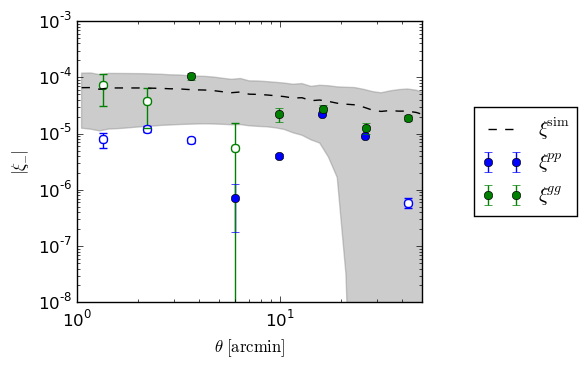}
 \caption[$\xi_{\pm}(\theta)$ Correlation Functions]{The $|\xi_{+}(\theta)|$ and $|\xi_{-}(\theta)|$ correlation functions for the $i^\prime$ band shape analysis as measured by \textsc{treecorr} are presented in the \textit{top} and \textit{bottom} panels respectively. The \textit{green} points show the galaxy auto correlations (the weak lensing signal,  $|\xi_{\pm}^{gg}(\theta)|$). The \textit{blue} points show the PSF auto correlations (pure systematics,  $|\xi_{\pm}^{pp}(\theta)|$) which are fully accounted for during the extraction of the galaxy shape measurements. The \textit{shaded} region and \textit{dashed} line shows the spread and mean of theoretical signals from the (SuperCLASS-like) \ih{supercluster} simulations describe in \cref{sec:radio_meas-simuclass-shear}. We provide the systematic check from \Cref{eq:syscheck} in \textit{red} for the $|\xi_{+}(\theta)|$ correlations (\textit{top panel}). At lower separations the recovered signal is significantly above the systematic checks. The absolute values for each are plotted, with unfilled markers indicating negative values.}
 \label{fig:gg_+-_corrs}
 \label{fig:corr_fns}
\end{figure}

For the SuperCLASS project, our expectations are the detection of the shear field and thus our requirements are not as stringent as other weak lensing studies \citep[e.g.][]{2016MNRAS.460.2245J,2015MNRAS.454.3500K}. We place requirements on the systematic contributions to $\xi_{\pm}(\theta)$, which must be significantly less than the expected signal. Hence, from \cref{eq:syst} we can form this requirement in terms of the bias parameters. For the PSF model analysis we are only concerned with placing requirements on the PSF leakage term:
\begin{equation}
|\alpha| < \left ( \frac{ \delta \xi_{\pm}(\theta) }{ \xi^{pp}_{\pm} (\theta) } \right )^{\frac{1}{2}}.
\end{equation}
For our detection purposes this provides the condition 
\begin{equation}\label{alpha_requ}
\alpha^{2} \xi^{pp}_{\pm} (\theta) <  \xi_{\pm}(\theta).
\end{equation}
We compute this PSF leakage in \cref{sec:optical_meas-shapes} using the galaxy shapes measurements presented there, and present the result in \cref{fig:gg_+-_corrs}.
\subsection{Shape measurement}
\label{sec:optical_meas-shapes}
With the image calibration and PSF model construction complete to suitable requirements the galaxy shapes can now be extracted from the data. We take advantage of ongoing weak lensing projects in optical studies. Specifically, we follow closely one of the shape fitting procedures implemented by \cite{2016MNRAS.460.2245J}, applying a maximum likelihood model-fitting algorithm, \textsc{im3shape} \citep{2013MNRAS.434.1604Z}, for estimating galaxy ellipticities.

\textsc{Im3shape} was shown to be a good method for galaxy shape estimation during the GREAT optical weak lensing challenges \citep{2012MNRAS.423.3163K,2012MNRAS.427.2711K,2015MNRAS.450.2963M}. Furthermore, \textsc{im3shape} allows for multi-epoch fitting by a simultaneous fit to exposures. This is an alternative approach to multi-exposure fitting by co-addition (or stacking) of exposures. As already mentioned in \cref{sec:optical_meas-psf}, we wish to minimise the resolution as much as possible to obtain the optimal number of unbiased shape estimates; co-addition would likely increase the minimum resolution available for any given source. Instead by the simultaneous fit to exposures, we are able to keep an individual PSF model per exposure.

For each source postage stamp, we also check for neighbouring sources in the cut-out area using \textsc{SExtractor} to identify source positions on a stamp-by-stamp basis. Flux from a neighbouring source can cause a bias in measured galaxy shapes and consequently on the measured shear. Even when considering an idealised (and isotropic) distribution of neighbours over an ensemble of galaxies, work by \cite{2018MNRAS.475.4524S} found a significant multiplicative $m$ shear bias arose. This showed strong dependence on distance to the nearest neighbour, and therefore the source density.

\begin{figure}
 \centering
 \includegraphics[width=0.5\textwidth,angle=0]{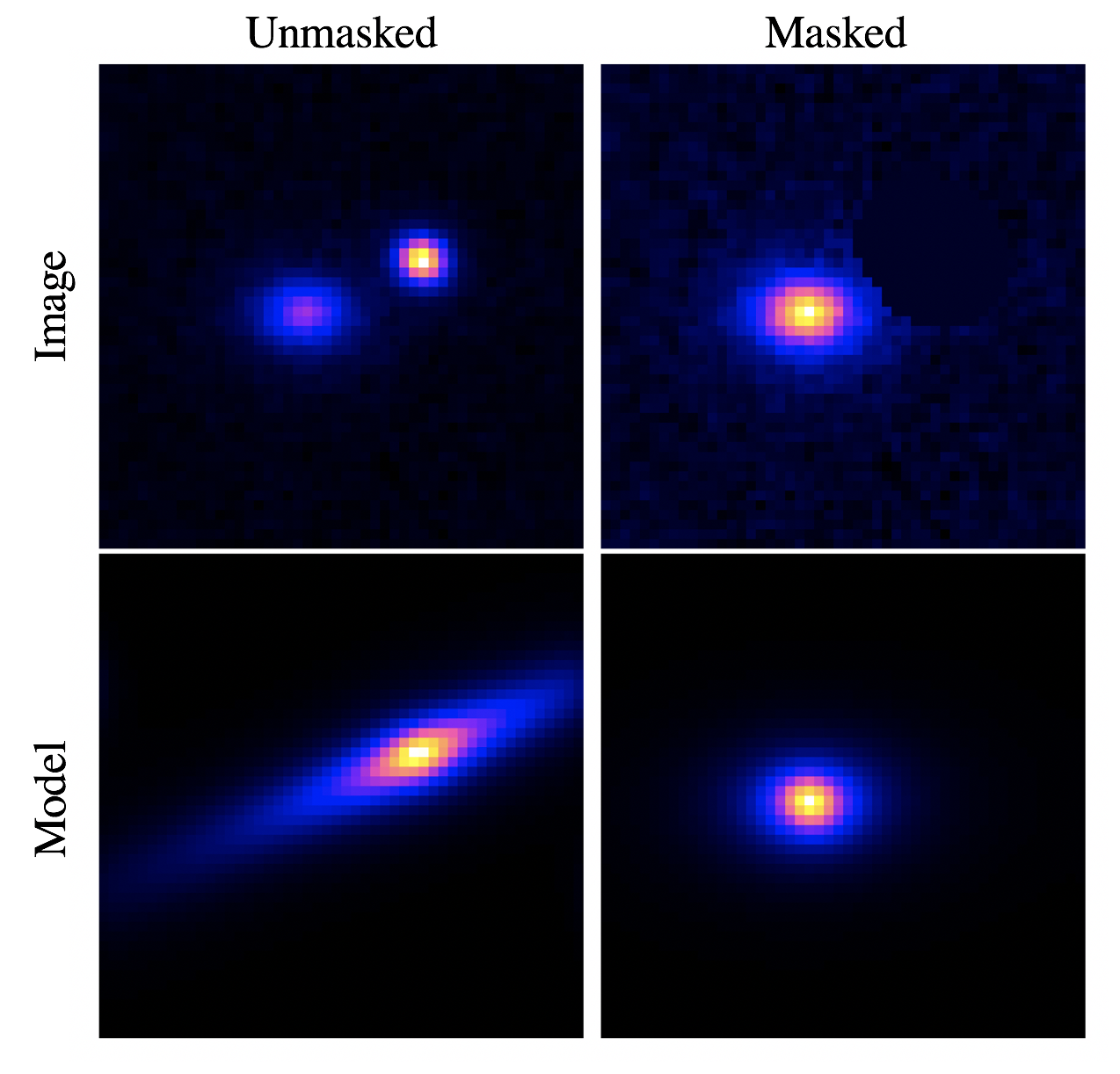}
 \caption[Source Fitting with Masked Neighbours]{Postage stamps from a single exposure for source ID $467466$ in the photometric catalogue, showing in the \emph{left} column the unmasked version and in the \emph{right} column with masking applied by multiplying with associated stamp weights. The \emph{top} row shows the image and the \emph{bottom} row the resulting \imshape fit for each case.}
 \label{fig:fit_masking}
\end{figure}

For minimising bias effects from neighbouring sources, we include the masking of the neighbours in the accompanying postage stamp weights. Pixels are associated with a given source using the shape information from a Gaussian elliptical fit measured in the postage stamp \textsc{SExtractor} source catalogue. Neighbours are searched for out to a distance of $1.5$ times the source FWHM, measured by \textsc{SExtractor}. The pixel weights are associated with the central source out to twice the measured FWHM; this ensures the central area remains unaffected by the neighbouring mask. An example of a masked \textsc{im3shape} fit is shown in \Cref{fig:fit_masking}. In this case the model fit is clearly improved when the neighbouring sources are masked. This is further indicated through the measured $\chi^{2}_{\mathrm{pixel}}$ which was found to be $38.99$ and $1.45$ for the un-masked and masked cases respectively.

For each galaxy position from the catalogue created in \cref{sec:optical_meas-catalogue}, $10\,{\rm arcsecond}\times 10\,{\rm arcsecond}$ image cut-outs from the present exposures and the appropriate PSF model images are constructed. We also run through the fitting procedure twice for each galaxy, since we found some issues with using the two-component summation of the bulge and disc fits. In many cases one of the components would be fit to a large negative amplitude which we wish to avoid. We instead fit first with a pure bulge (de Vaucouleurs) and then with a pure disc (exponential) model, fixing the other component to zero in each case. In the final catalogue we keep the fit with the closest $\chi^{2}_{\mathrm{pixel}}$ to 1 for each galaxy.

\subsection{Shear correlation function}
\label{sec:optical_meas-shear_correlation}
In order to assess the quality of the shape measurements discussed above, we make use of the real-space two-point correlation functions of galaxy shapes.

We calculate the two-point correlation function, corrected for sky curvature, from the shape catalogue produced in \cref{sec:optical_meas-shapes}, using the freely available code \textsc{treecorr}\footnote{https://github.com/rmjarvis/TreeCorr} \citep{2004MNRAS.352..338J,2015ascl.soft08007J} which has been specifically designed for use with cosmology and weak lensing studies. The shear two-point correlation function is parametrised by $\xi_{+}^{gg}(\theta)$ and $\xi_{-}^{gg}(\theta)$ ($gg$: galaxy-galaxy) which are shown in \Cref{fig:gg_+-_corrs}. In each of these plots we also include the expected signal derived from the $N$-body \ih{supercluster} simulations (see \cref{sec:radio_meas-simuclass-shear}). The shear two-point correlation functions are calculated with \textsc{treecorr} for the full range of uncertainty given by the simulation realisations. The measured signal from the Suprime-Cam data, also displayed for both $\xi_{+}^{gg}(\theta)$ and $\xi_{-}^{gg}(\theta)$, lies comfortably in the shaded region (which represents the spread in correlation functions from the different clusters in the simulations), signifying that the data agrees with the theoretical signal expected from the \ih{supercluster} simulations.

In previous studies a number of tests have been used to assess the quality of data by searching for signals which would be zero without the presence of systematic errors in the data. We perform a direct comparison of the galaxy correlation function $\xi_{+}^{gg}(\theta)$, to the galaxy-PSF shape cross-correlations $\xi^{gp}_{+}$:
\begin{equation}
\xi^{gp}_{+} = \langle e^{*} (r) \rangle \langle e_{\mathrm{psf}} (r+\theta) \rangle,
\end{equation}
where $e_{\mathrm{psf}}(r)$ is the shape of the PSF surface brightness model at position $r$.

This cross-correlation test will expose any PSF leakage onto the galaxy shapes, potentially a major source of shape bias. We parametrise this test in terms of the PSF leakage parameter $\alpha$ in order to relate back to our noted requirements in \cref{sec:optical_meas-psf_diagnostics}, where for our detection purposes, $\alpha^{2} \xi^{pp}_{\pm} (\theta) < \xi_{\pm}^{gg}(\theta)$. We can re-write \cref{eq:bias_form} in terms of $\xi^{gp}_{+}$, to solve for $\alpha$,
\begin{equation}\label{eq:syscheck}
\alpha = \frac{\xi^{gp}_{+} - \langle e_{\mathrm{gal}} \rangle^{*} \langle e_{\mathrm{PSF}} \rangle } { \xi^{pp}_{+} - | \langle e_{\mathrm{PSF}} \rangle |^{2} },
\end{equation}
as shown in \cite{2016MNRAS.460.2245J}. We make a comparison of the detected signal in the correlation function to this statistic in \cref{fig:gg_+-_corrs} where we can see that the lensing signal inferred from the $\xi^{gg}_{+}$ is significantly above the systematic statistic $\alpha^{2} \xi^{pp}_{+}$, at least up to scales of $\sim10\:\mathrm{arcmin}$. 
A `pure' systematic signal is shown by the PSF shape auto-correlations, $\xi^{pp}_{+}$ for comparison. We also show the $\xi^{gg}_{-}$ statistic, which again is consistent with expectations from the simulated \ih{supercluster} regions.
\subsection{Redshift distributions}
\label{sec:optical_meas-redshifts}
For a full description of the initial redshift analysis of the optical data, we refer to Paper II. Redshifts are derived using the $BVr^\prime i^\prime z^\prime Y$ photometry from Subaru, plus IR data from the \emph{Spitzer} telescope, and by using the EAZY \citep{2008ApJ...686.1503B} template fitting code. The lack of full coverage of the field in the $z^\prime$ band, and lack of any coverage in the $u$ band, means there are significant degeneracies between low-redshift $z<0.5$ and high-redshift $z>2$ templates, making many of the redshifts unreliable, particularly in the range $0.2 < z < 0.8$. Therefore, other than isolating the cluster members we do not use the photometric redshifts for these sources in our weak lensing analysis, choosing instead to perform a 2D cosmic shear analysis, without tomographic binning, on the catalogue described above in \cref{sec:optical_meas-catalogue}.
\subsection{Shear maps}
\begin{figure}
\includegraphics[width=0.5\textwidth]{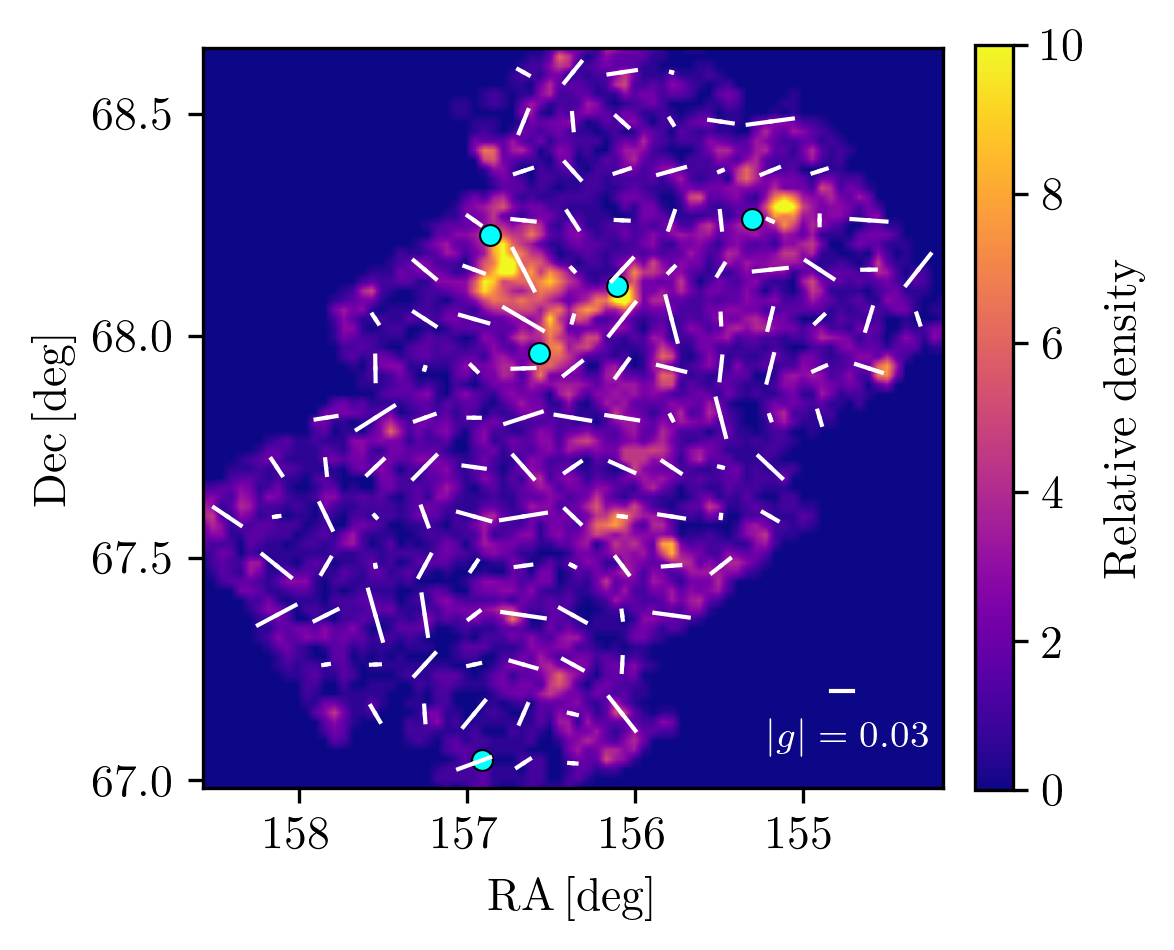}\\
\includegraphics[width=0.5\textwidth]{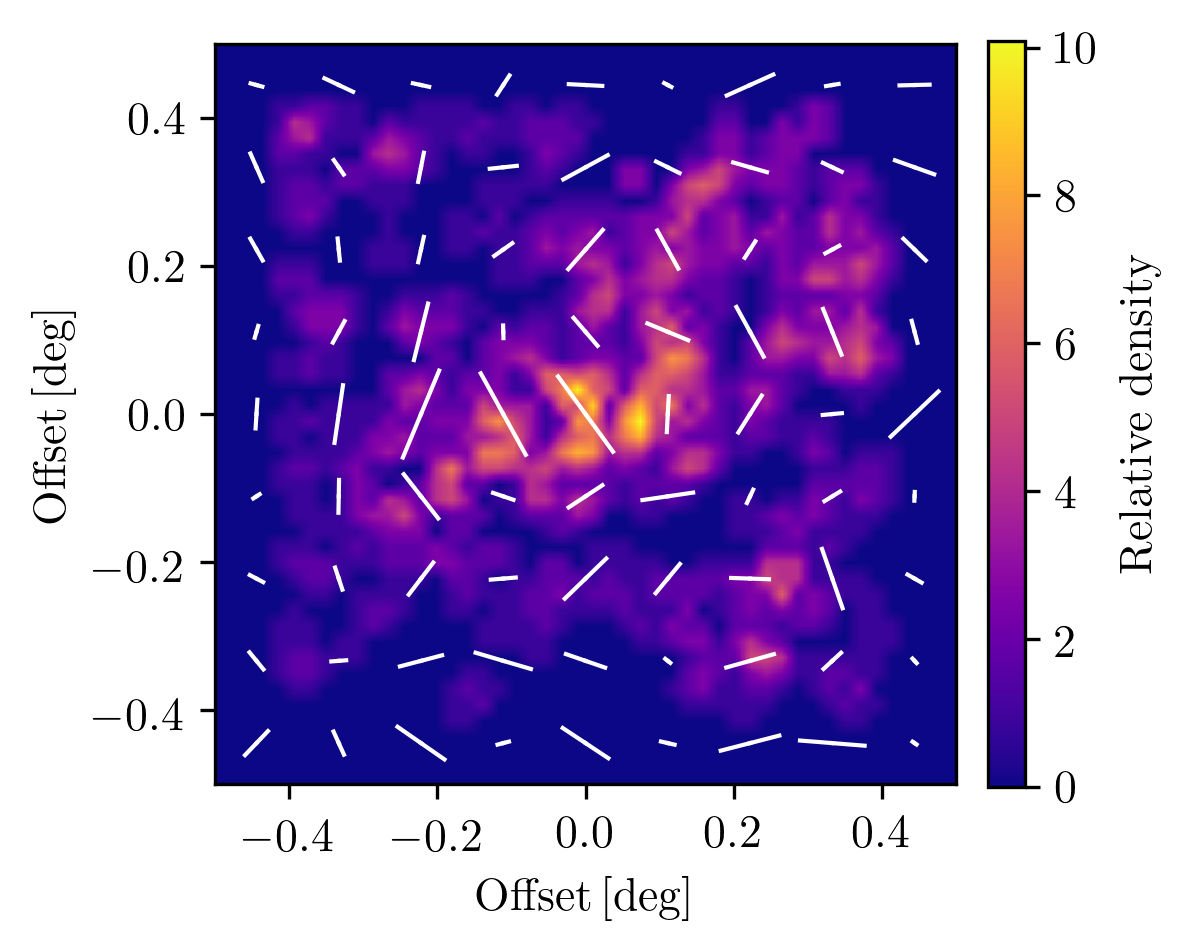}
\caption{Maps of relative galaxy density (colour scale) and measured shear signal (ticks) from the SuperCLASS optical data (\emph{upper}, with Abell cluster locations also shown as cyan circles) and a simulated data set with the same noise properties and a shear signal given by a comparable \ih{supercluster} region (\emph{lower}). Note the different spatial extents and aspect ratios between the two plots.}
\label{fig:shear_maps}
\end{figure}

In addition to the two-point shear statistics described in \cref{sec:optical_meas-shear_correlation} we also create maps of the shear measured from the optical shapes in $6.67\,$arcmin pixels, as shown in the upper panel of \cref{fig:shear_maps}, plotted on top of the relative density of galaxies in the photometric data set which have $z_{\rm phot}=0.2\pm0.08$ and $i^\prime<22.5$ (see Paper II, section 4.3.2). Also plotted for reference are the locations of the Abell clusters forming the \ih{supercluster} field. We validate this pipeline by also running it on a simulation. In the lower panel of \cref{fig:shear_maps} we show the shear signal in a comparable simulation of the SuperCLASS optical observations. This simulation has a shear signal calculated as described in \cref{sec:radio_meas-simuclass-shear} for one of the \ih{supercluster} regions. The distribution of foreground sources is given by Poisson sampling the corresponding convergence map and ensuring number densities match between simulation and data. Shape noise is included by randomly rotating the shapes of sources in the real optical catalogue and placing them in a uniform distribution across the map. We note that the lack of visual correspondence between the cluster overdensity field and the shear field is likely due to a number of reasons, including shot noise due to intrinsic galaxy shapes, residual shear measurement systematics, and foreground galaxies in the lensing sample. For a rigorous assessment of the significance with which we detect a shear signal, see \cref{sec:power_spectra} where we calculate the two-point statistics of the measured shear field. 
\section{Radio shape measurement}
\label{sec:radio_meas}
As discussed in \cref{sec:introduction}, SuperCLASS is the first survey to be designed with the express purpose of measuring a weak lensing signal in the radio. The most basic building block of a weak lensing cosmic shear measurement is a catalogue of shapes of distant galaxies, from which the shear may be inferred. Highly precise and accurate shape measurement methods have been developed over a number of years for CCD images \citep[e.g. the compilation of][]{2015MNRAS.450.2963M}, but for radio interferometer data, the situation is less well developed. \ih{A number of approaches have been proposed \citep{2018MNRAS.476.2053R, 2019MNRAS.482.1096R} but have only been tested on simulations, and not real data. Conversely the method used in \cite{2004ApJ...617..794C} was applied directly to data, but not simulations (still less simulations of the more recent data sets which are far greater in volume).} Here we detail the identification of sources in our data suitable for weak lensing shape measurement, and our method for recovering the shapes of sources in the image plane (which we call SuperCALS and describe below in \cref{sec:radio_meas-supercals}).
\subsection{Weak lensing catalogue}
\label{sec:radio_meas-catalogue}
For a full description of the SuperCLASS catalogue generation, we refer to Paper I, Section 4. In short, from images produced with the \clean algorithm, we run the \pybdsf \citep{2015ascl.soft02007M} source finding algorithm, which both estimates the noise in the image and finds sources by fitting multiple Gaussian profiles. The weak lensing catalogue is a subset of this full catalogue, with the selection cuts described here. All the sources described are in the DR1 region defined above.
\subsubsection{VLA catalogue}
\label{sec:radio_meas-cataloguep-VLA}
The VLA weak lensing catalogue is created from the VLA DR1 catalogue, whose creation is described in Paper I. From this catalogue we then select sources which are resolved according to the PyBDSF output columns:
\begin{align}
    &\mathtt{DC\_Min} > 0 \nonumber\\
    \mathtt{and} \,\, & \mathtt{DC\_Maj} > 0 \nonumber\\
    \mathtt{and} \,\, & \mathtt{Maj} > \mathtt{BMAJ}, \nonumber
\end{align}
where \texttt{DC\_Min} and \texttt{DC\_Maj} are the PyBDSF deconvolved major and minor axes, \texttt{Maj} is the PyBDSF convolved major axis and \texttt{BMAJ} is the restoring beam major axis, used in the creation of the deconvolved image. We also impose a cut to ensure the sources have high signal-to-noise ratio:
\begin{align}
    &\mathtt{Total\_flux} > 50\,\mu\mathrm{Jy}, \nonumber
\end{align}
where \texttt{Total\_flux} is again measured by PyBDSF, corresponding to a typical SNR cut of SNR$\,>7$.

We also make a cut to keep only sources which are visually consistent with having `simple morphology' -- which we use as a proxy for removing contaminating Active Galactic Nuclei (AGN) sources. AGN are expected to be the source of a significant fraction of the emission in the radio sky at L-band frequencies, and themselves have a rich taxonomy of spectral and morphological sub-classes. Weak lensing shape measurement using simple models with elliptical isophotes, as used here, relies on there being a small `model bias' between such models and the true galaxy flux \citep[see e.g.][]{2010MNRAS.404..458V}, but this bias will be large when fitting a simple model to a complex AGN, meaning we choose to discard them for shape measurement.

Here we assume that all sources are heavily dominated by either i) synchrotron emission from star-forming regions, or ii) emission from jets and hot spots associated with AGN. Sources in the class i) are identified by having visually simple morphology, whilst sources in class ii) are expected to have more complex morphologies consisting of multiple components and flux peaks. This classification was performed visually by multiple members of the SuperCLASS collaboration, using tools from the Zooniverse \citep[e.g.][and Paper I Section 4.4]{2012amld.book..213F}, with postage stamp images of each source to be classified from \emerlin, VLA and Subaru data presented next to each other. Users are asked to classify the sources as simple or complex morphology. Sources are then included in the weak lensing catalogue when a majority of users classify the source as having simple morphology. More sophisticated ways of classifying radio sources between AGN and star-forming galaxy categories exist (such as those making use of radio-infra-red correlations), but here we use this simple criterion with the goal of maximising the number of sources available for shape measurement rather than losing information due to e.g. lack of infra-red detections.

\ih{Some of these sources may in fact have emission dominated by a Radio Quiet AGN (RQ-AGN) component \citep[see e.g.][]{2015MNRAS.452.1263P,2018MNRAS.481.4548P}. Cleanly separating these populations is usually done by combining L-band radio data with $24\,\mu$m and X-ray data, which we do not have available for this field. However, when plotting the joint distribution of best fitting \sersic index and radius for the sources, we do not see any excess of sources which have a high \sersic index and small radius, as may be expected for contaminating RQ-AGN.}

\ih{When these cuts are applied and the SuperCALS shape measurement method is applied (as in \cref{sec:radio_meas-supercals}) we are able to measure 440 shapes in the DR1 region, corresponding to a radio weak lensing source number density of $n_{\rm gal}^{\rm R} \approx 0.47\:\mathrm{arcmin}^{-2}$.}
\subsubsection{\emerlin catalogue}
\label{sec:radio_meas-catalogue-emerlin}
The \emerlin weak lensing catalogue is created by cross-matching the \emerlin DR1 catalogue (as described in Paper I) with the VLA DR1 weak lensing catalogue. This leads to a total of 56 sources available for shape measurement in the $\sim0.26\degsq$ area, a source density of 0.06 arcmin$^{-2}$. The \emerlin catalogue is defined in this conservative way in order to avoid spurious detections in the weak lensing catalogue. The nature of the \emerlin PSF and noise correlations induced by the sparse $uv$-plane coverage mean that high fractions of detections from running the PyBDSF source finding code were identified as false detections -- with similar numbers being found in the \emph{negative} image (i.e. the map multiplied by $-1$, which should contain no real sources with negative flux). We therefore choose to include \emerlin information as additional shape information, where available, for the VLA catalogue in separate columns in the weak lensing catalogue which are not used in the main science analysis in \cref{sec:power_spectra}.
\subsection{Simulation pipeline}
\label{sec:radio_meas-simuclass}
The process by which radiation falling on to a radio telescope is turned into estimated measurements of the cosmic shear signal along a given line of sight is a highly complicated, non-linear process. Many aspects of this process are potentially capable of introducing a spurious shear signal into our data which may be mistaken for the true signal. In order to evaluate and quantify these systematic error effects we have constructed a simulation pipeline, referred to as SimuCLASS, which seeks to replicate as far as possible the full forward model applied to create a shear catalogue from the sky brightness distribution: performing exactly the same operations on the simulated data as are applied to the real data. We then inject sky models with known shear properties into the pipeline and attempt to recover the signal using our shear measurement method. This allows us to both calibrate and validate the method. The SimuCLASS pipeline comprises four main parts, with the simulated visibility data then being fed into the imaging and source finding pipeline described in \cref{sec:radio_meas-catalogue} and Paper I, before analysis with the shape measurement method described in \cref{sec:radio_meas-supercals}.
\subsubsection{Population Model}
\label{sec:radio_meas-simuclass-trecs}
For our population model we use the Tiered-Radio Extragalactic Continuum Simulation (T-RECS) of \cite{bonaldi-trecs}. This is 
a new simulation of the radio sky in continuum, which reproduces the most recent compilation of data in terms of number counts, luminosity functions and redshift distributions over the 150\,MHz -- 20\,GHz range. Of the observational parameters modelled in T-RECS, those relevant for this work are: the position on the sky, the integrated flux at 1.4\,GHz, the source class (either Active Galactic Nuclei, AGNs, or Star-Forming Galaxies, SFGs) and the source size and shape. 

Starting from a dark matter-only $N$-body simulation of $800 \,h^{-1} \,$Mpc$^3$, haloes are found and grouped together down to a mass of $\sim10^8\,h^{-1}\,M_{\odot}$ and light cones constructed in a $5\times5$ deg sky area out to a redshift of $z=8$. Abundance matching methods are then used to assign galaxies to haloes, thus giving realistic clustering properties to the galaxies of each population. 

The shape and size of AGNs is modelled in T-RECS in terms of a largest angular size and distance between the hot spots, in a way that reproduces the observed correlation with both flat/steep spectrum and FRI/FRII classifications. 

The shape and size of SFGs is modelled as an exponential intensity profile: 
\begin{equation}
I(r) = I_{0} \exp\left( -r/r_0 \right),\label{expo}
\end{equation}
where $I_{0}$ is a flux normalisation, and $r_0$ is the scale radius. The sources are given an intrinsic ellipticity $|\eint|$ from the distribution found in radio VLA observations of the COSMOS field as in \cite{2016MNRAS.463.3339T}:
\begin{equation}
    P(|\eint|) = |\eint| \left[\cos\left(\frac{\pi |\eint|}{2}\right)\right]^{2} \exp\left[-\left(\frac{2|\eint|}{B}\right)^{C}\right],
\label{eqn:eint}
\end{equation}
with the best-fitting parameters $B = 0.113\pm 0.041$ and $C = 0.303\pm 0.058$, giving a shape noise dispersion of $\sigma_\eint$ = $0.29$ (per ellipticity component). 
\subsubsection{Shear Signals}
\label{sec:radio_meas-simuclass-shear}
In order to model the weak lensing shear expected in the SuperCLASS field, we make use of a suite of $N$-body simulations to model the expected signal. Fully described in \cite{2018MNRAS.474.3173P}, this simulation consists of $2520^{3}$ dark matter particles of $5.43\times10^{10}\,h^{-1}M_{\odot}$ evolved from redshift $z=127$ to $z=0$. \ih{Superclusters} are then identified in the simulation which have similar properties to those expected in the SuperCLASS field -- regions with five cluster members in the $z=0.24$ output snapshot which are linked with a friends-of-friends algorithm with linking length $l=8 \,h^{-1}$Mpc. This identifies 61 \ih{supercluster} regions which are then re-simulated, along with 60 randomly chosen regions, at a higher resolution as part of the MACSIS project \citep{2017MNRAS.465..213B, 2017MNRAS.465.3361H}. This involved dark matter only (particle mass $5.2\times10^{9}\,h^{-1}M_{\odot}$) and full hydrodynamical (initial gas particle mass $8.0\times10^{8}\,h^{-1}M_{\odot}$) simulations. The \ih{supercluster} regions allow us to predict the measurable enhancement of the matter power spectrum and, using techniques developed as part of the SUNGLASS pipeline \citep{2011MNRAS.414.2235K}, the enhancement of the weak lensing shear power spectrum, which may be expected in the SuperCLASS region over a randomly chosen field. This enhancement is found to be a factor $\sim2$ with variations between 1 and 3. The mean measured shear power spectum from the 61 \ih{supercluster} regions is used to represent theoretical expectations for the shear power spectra measured in  \cref{sec:power_spectra}.
\subsubsection{Sky Models}
\label{sec:radio_meas-simuclass-sky}
The simulation pipeline creates models of sky emission using the GalSim galaxy image simulation toolkit \citep{2015A&C....10..121R}, a fast and accurate package for creating simulated galaxy intensity profiles. We place sources with properties in accordance with a simulated catalogue from the T-RECS simulation (see \cref{sec:radio_meas-simuclass-trecs}), from which we take the sky position (with respect to the pointing centre), source type, flux, size, and intrinsic source shape. Each source is also sheared with the correct amount of cosmic shear for its sky position according to the shear signal simulation (\cref{sec:radio_meas-simuclass-shear}).

We use the same sky models to model the entire observed radio bandwidth (i.e. assume flat spectral indices across the sources), which is not strictly correct. However, spectral imaging is not performed on the real data, meaning morphological information available is averaged over the bandwidth into a single image in the same way, and the shear signal we are looking for is independent of wavelength.

Sources of extragalactic radio emission at the $\sim 1\,\mathrm{GHz}$ frequencies and $\mu\,\mathrm{Jy}$ fluxes considered here are expected to be a mixture of SFGs and AGN. Individual SFGs are expected to have (on average) simple emission profiles with elliptical isophotes, and it is these objects which we seek to measure the shape of in order to form an estimator for weak lensing shear. Resolved AGN are expected to have significantly more complicated morphologies, meaning we do not consider them for shape measurement, but they can still be included in simulations in order to quantify effects from e.g. their un-deconvolved sidelobe noise.

For the purposes of DR1 we are using shapes only from relatively high signal-to-noise objects in a relatively small field, meaning we choose not to include AGN in the simulations used for the SuperCALS calibration in \cref{sec:radio_meas-supercals}.

Star-forming galaxies are drawn with the exponential intensity profile (\cref{expo}) and intrinsic ellipticity (\cref{eqn:eint}) included in the T-RECS catalogue. We draw the values for the galaxy position angle $\theta$ from a uniform distribution. Finally, a shear $\gamma$ from the expected (from the $N$-body simulation described in \cref{sec:radio_meas-simuclass-shear}) weak lensing signal at the sky location of the galaxy is added, and the flux is added over the correct number of pixels in the full image out to a maximum radius $r_{995}$ such that $99.5\%$ of the flux is placed in the image.
\subsubsection{Interferometer model}
\label{sec:radio_meas-simuclass-smtool}
We then model the observation of these simulated skies by the \VLA and \emerlin radio telescopes. Schematically, radio interferometers turn a real, two-dimensional sky brightness distribution $I(l,m)$ into a three-dimensional data set of complex-valued \emph{visibilities} \citep[see e.g.][]{2017isra.book.....T}:
\begin{equation}
\label{eqn:int_transform}
\tilde{V}(u, v, w) = \mathbf{MFCA} \, I(l,m) + N.
\end{equation}
Here, the operators $\mathbf{MFCA}$ represent different components of the forward model. $\mathbf{A}$ is the primary beam, giving the response to the sky of an individual element within the interferometer.  $\mathbf{C}$ is the $w$-term, representing the projection from the two-dimensional sky to the three-dimensional space in which the antennas exist (i.e. due to the Earth's curvature). $\mathbf{F}$ is the Fourier transform operation, and $\mathbf{M}$ is the masking function, representing the sampling of the full visibility space by the finite set of observations present in the data. Each sample represents an integration over a small time and frequency interval at a spatial scale represented by the baseline between the pairs of antennas in the array. $N$ is a noise term, typically taken to be Gaussian distributed and uncorrelated between visibility points.

\begin{figure*}
\includegraphics[width=0.5\textwidth]{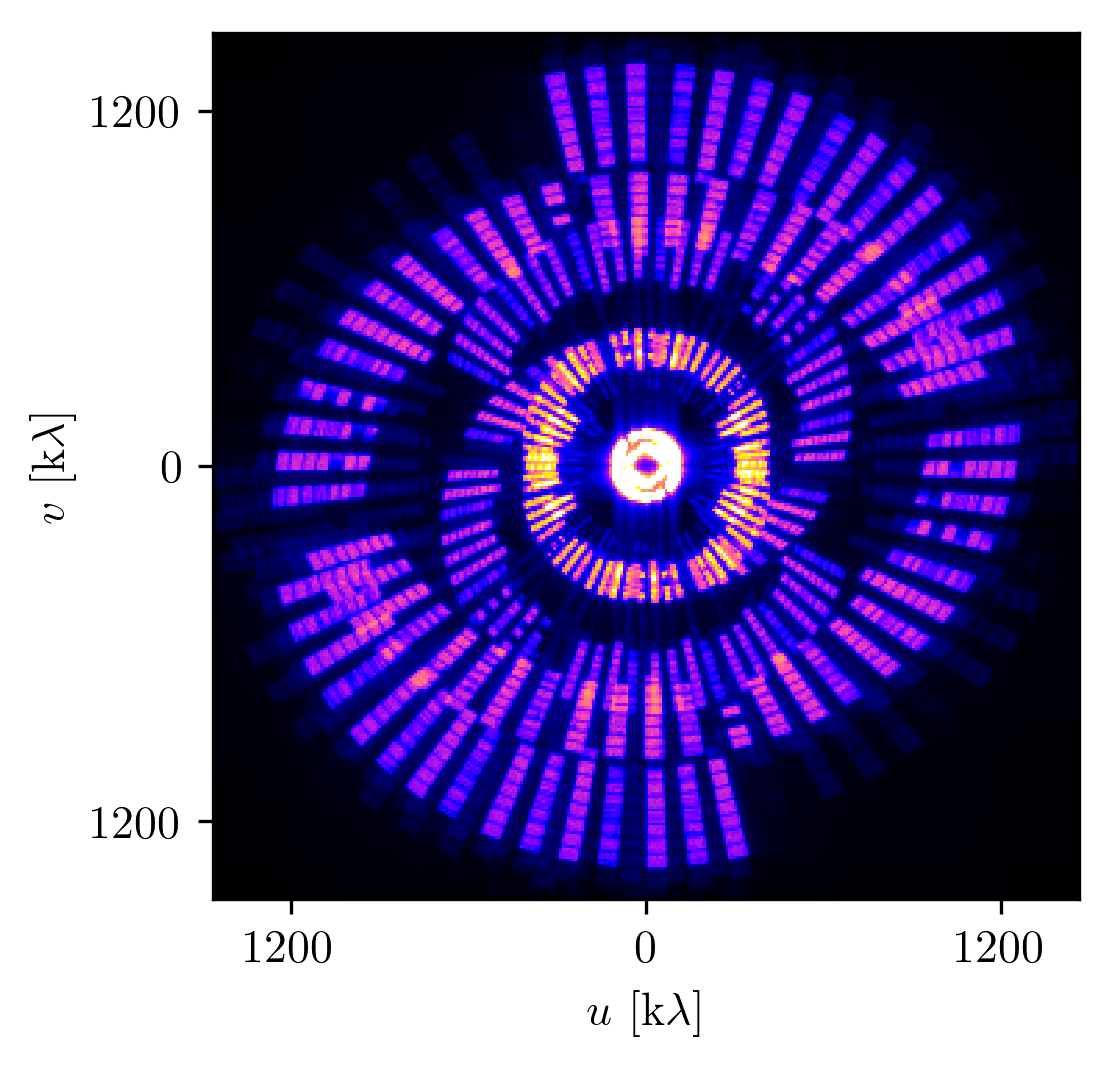}\includegraphics[width=0.5\textwidth]{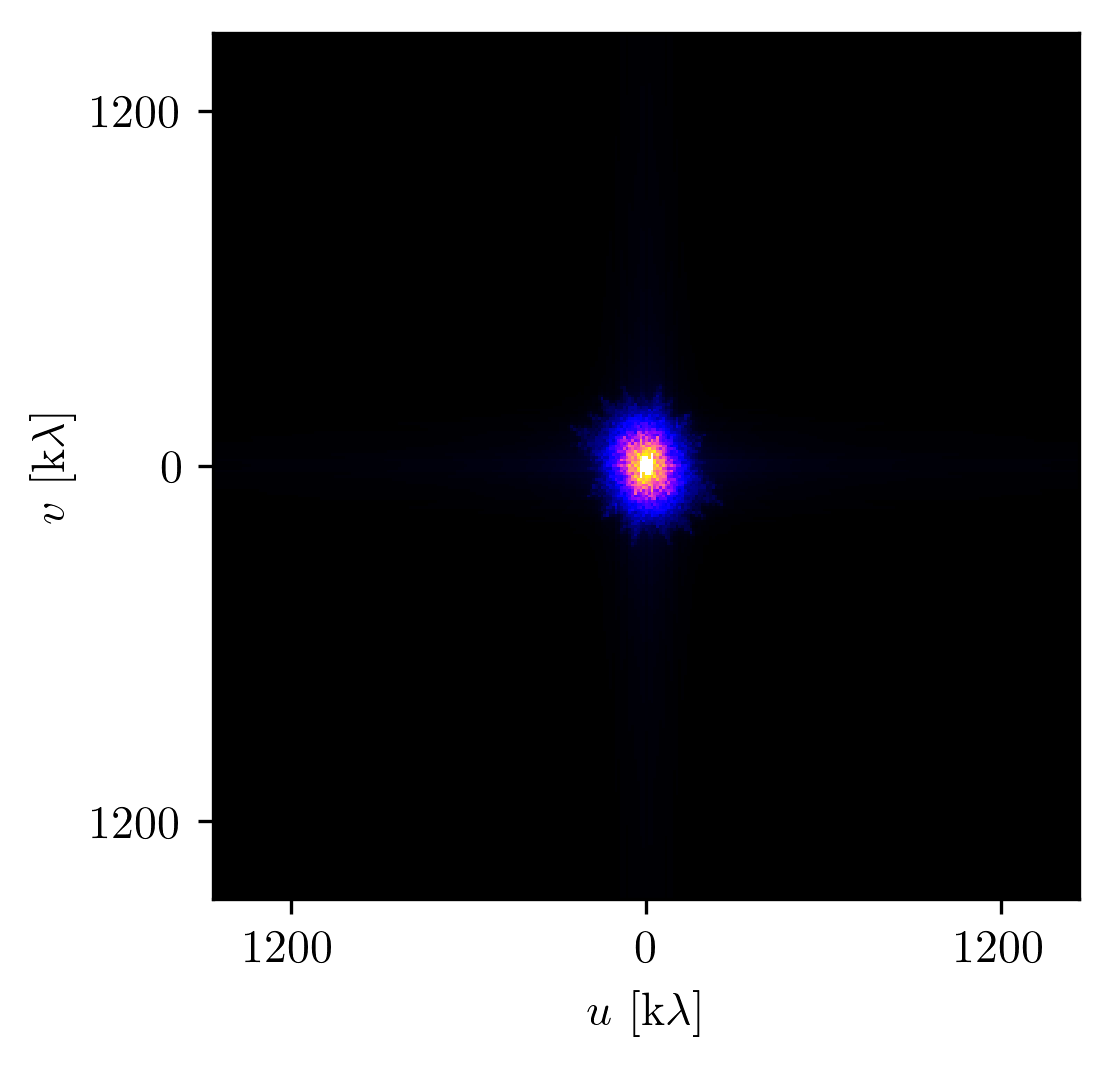}
\caption{$uv$ plane coverage of the J28 pointing for (\emph{left}) \emerlin and (\emph{right}) VLA on Fourier scales in units of kilo-wavelengths. Colours show the density of visibility data, with bright regions having higher density, and hence lower noise on the corresponding Fourier scales. Black areas represent gaps in the $uv$ coverage where no telescope baselines exist to make a measurement.}
\label{fig:uvcoverage}
\end{figure*}

Visibilities are an incomplete sampling of a 3D integral transform (Fourier transform plus the $w$-term) of the sky brightness distribution. \Cref{fig:uvcoverage} shows the inverse-noise weighted density of the Fourier $uv$ plane sampling in one pointing of the VLA and \emerlin observations from SuperCLASS. Transforming this coverage back into the image plane provides the PSF (usually referred to as the `dirty beam') which is convolved with the sky. This beam is set by the time and frequency samplings and the spatial distribution of antennas in the telescope, all of which are known to a high degree of precision. However, the missing information from unsampled parts of the Fourier plane leads to significant PSF sidelobes which extend across the entire sky. These sidelobes mix information from multiple sources, making shape measurement of individual sources highly challenging and potentially requiring simultaneous shape measurement across many sources as described in \cite{2019MNRAS.482.1096R}.
The visibilities here are created using the simulation tools in the \casa radio astronomy package \citep{2007ASPC..376..127M}. In order that the dirty beam PSF of our simulations matches as closely as possible that of the real data, we make use of the $uv$ coverage available on each pointing centre in the real data, after losses due to telescope outages and Radio Frequency Interference (RFI) removal. This means we can create simulated images which have exactly the same dirty beam PSF as the real ones.

From the real data pipeline (see Paper I, Section 3) we obtain single \casa measurement sets corresponding to the full set of observations for each of the SuperCLASS pointings, separately for the \emerlin and VLA telescopes. The \casa simulator tool is then used as \texttt{sm.openfromms} and \texttt{sm.predict} in order to project the simulated sky brightness distribution onto the visibilities. This prediction consists of the full three dimensional transform, including the $w$-term.

Noise is added as uncorrelated Gaussian random variates to each visibility point, with a variance of $0.4\,$Jy in order to match the image-plane noise levels. Here, we assume no calibration errors (such as residual phase errors) are present in the data, seeking only to assess the impact of noisy interferometric imaging observation and imaging reconstruction, but such effects should be included in future.

\subsubsection{Imaging deconvolution}
\label{sec:radio_meas-simuclass-imaging}
Imaging of each simulation is carried out using the same algorithms and settings as for the real data, to ensure shape measurement artefacts are correctly mimicked. Simulated VLA data are imaged (as in \cref{sec:survey-radio-VLA}) using the \casa ~{\tt tclean} task, with multi-scale \clean and Briggs weighting, while \emerlin data are imaged (as in \cref{sec:survey-radio-emerlin}) using WSCLEAN, using natural weighting. As no sources brighter than $500\,\mu$Jy are included in the sky model, there is no need to account for the effects of such sources via peeling or self-calibration, but this means our simulations also do not include detrimental effects from incompletely removed bright confusing sources.
\subsection{Shape measurement with SuperCALS}
\label{sec:radio_meas-supercals}
Although images recovered from radio interferometer data have been shown to contain useful morphological information, galaxy ellipticity measurements from such images are typically highly biased \citep{2015aska.confE..30P}. Here, we implement an image-plane shape measurement method on the images produced from our data which involves a step to calibrate these biases, referring to this method as SuperCALS (Super Calibration of All Lensed Sources). The philosophy of this method originates in the many optical shear measurement methods which rely on a step in which simulated sources are injected into the noise environment of the real data \cite[e.g. MetaCal, ][]{2017ApJ...841...24S}. Even if a method produces a shear measurement which is biased, if we can produce simulations which are sufficiently representative of the real data (specifically in that the method has the same bias with respect to the simulations as the real data) then we may inject data with known shear signals into the method, recover the biased results and then use these bias values to calibrate our shear estimates in the real data images. This is reliant on the biases created being relatively small, i.e. still well-modelled by first order shear transformations of galaxy profiles with elliptical isophotes. The results in \cref{sec:radio_meas-supercals-sims} show that this appears to be the case for the majority of the sources of interest.
\subsubsection{Method}
\label{sec:radio_meas-supercals-method}
We first make an image from the VLA or \emerlin data using the \clean algorithm, as described in \cref{sec:radio_meas-simuclass-imaging} (and in Paper I, Section 3). The \clean algorithm creates a model of the original sky brightness distribution by iteratively deconvolving the dirty beam PSF. This process is highly non-linear, and has noise properties which are hard to estimate. The outputs from this process are:
\begin{itemize}
\item The dirty beam PSF: the image plane representation of the $uv$ coverage. This is the PSF which is convolved with the sky brightness distribution in the observation, and is precisely known and highly deterministic, but has significant sidelobes extending across the entire field, meaning confusion is a problem for all sources.
\item The `dirty image', consisting of the transformation of the data into the image plane, giving the sky emission convolved with the dirty beam PSF.
\item The `model image', consisting of a set of source models (typically Dirac $\delta$-functions) of varying brightness and sky location, representing the deconvolved sky brightness emission as determined via the \clean algorithm.
\item The `\clean image', consisting of the addition of the flux from the residual image (see below), plus that from the convolution of the Dirac $\delta$-function model image with a Gaussian `\clean beam' representative of the central lobe of the full dirty beam (i.e. with no sidelobes).  Note that in the \casa \clean task used for the VLA data, both the residual image and the Dirac $\delta$-function model image is convolved with the \clean beam, leading to additional noise correlations (at least when the restoring beam is manually fixed, as it is here\footnote{See note on {\tt restoringbeam} argument at \url{https://casa.nrao.edu/docs/taskref/tclean-task.html}}).
\item The `residual image', consisting of the remaining flux after the set of Dirac $\delta$-functions in the model image is convolved with the dirty beam and subtracted from the dirty image.
\end{itemize}
Rather than measuring shear directly from the \clean image, we rather use the output residual image to model the `noise' on a shape measurement in our data at a given sky location, both random (but correlated in the image plane) thermal noise and systematic noise from un-deconvolved or incorrectly deconvolved sidelobes from other sources. We then use this model to correct the shear measurement from the \clean image.

We first perform an initial estimate of source shapes by running the \imshape code \citep{2013MNRAS.434.1604Z} on the images. As described in relation to the optical data in \cref{sec:optical_meas-shapes}, \imshape performs a maximum likelihood fit of elliptical \sersic intensity profiles convolved with a PSF model to image plane data, and has been shown to peform at a high level of precision and accuracy when recovering cosmic shear signals from optical images \citep{2015MNRAS.450.2963M}. For this first run, we use the \clean image and \clean beam to estimate the source shapes. \imshape is run three times in different modes: once fitting a single Gaussian profile to each source, once fitting a \sersic profile with free index, and once fitting a joint bulge (\sersic $n=4$) plus disk (\sersic $n=1$) profile. The best-fitting of these three runs is then chosen, with subsequent \imshape runs for a given source retaining the fixed combination of number of components and \sersic indices for their fitted profiles. We have found this approach allows a good fit to be found for nearly all of the weak lensing catalogue sources, with 24\% having Gaussian fits, 67\% having free-\sersic fits and the remainder having bulge plus disk fits.

For each source, we then inject model sources with the same size and flux properties, but known ellipticity, into the residual image. We then perform a `ring test' to remove the effect of shape noise \citep{2007AJ....133.1763N} and create a model of the bias on the \imshape measured ellipticity at this sky position as a 2D surface. For each source position on the sky (labelled $k$) we find the bias between measured and input ellipticity component $e_1$ as a function of both input $e_1$ and $e_2$:
\begin{equation}
    b_{k}(e^{\rm inp}_1, e^{\rm inp}_2) = e^{\rm obs}_1 - e^{\rm inp}_1.
    \label{eqn:supercals_bias_surface}
\end{equation}
i.e, $b_k$ is a 2D surface with height $e^{\rm obs}_1 - e^{\rm inp}_1$ at location $(e^{\rm inp}_1, e^{\rm inp}_2)$. We construct a similar surface with the heights given by $e^{\rm obs}_2 - e^{\rm inp}_2$. Theses surfaces are interpolated between injected simulations using second-order 2D polynomials. Specifically, we run \imshape, fixed to the combination of \sersic profiles already determined for this source, on a total of
\ih{33 different injected sources,with $e_i = \lbrace0, \pm 0.2375, \pm 0.475, \pm 7125, \pm 0.95 \rbrace$.} Our estimate of the true shape for a given source $k$ is then its initial \imshape estimate  from the \clean image, corrected by the bias surface calculated using the residual image and simulations, evaluated at the measured uncorrected ellipticity:
\begin{equation}
    e^{\rm SC}_{1,k} = e^{\rm obs}_{1,k} - b_k(e^{\rm obs}_{1,k}, e^{\rm obs}_{2,k}),
\end{equation}
and similar for $e_2$. In the weak lensing catalogue, we include both the uncorrected source ellipticity from \imshape ($e^{\rm obs}_{1,k}, e^{\rm obs}_{2,k}$), and the ellipticity estimated by the SuperCALS method ($e^{\rm SC}_{1,k}, e^{\rm SC}_{2,k}$). We refer to this calibration as \emph{source-level} calibration.
\subsubsection{Performance on simulations}
\label{sec:radio_meas-supercals-sims}
We evaluate the performance of SuperCALS on simulated images of the VLA and \emerlin data created using the SimuCLASS pipeline described in \cref{sec:radio_meas-simuclass}. We simulated pointings of the VLA data, injecting only star-forming galaxy sources with total flux $>50\,\mu$Jy into the sky model, giving $\sim400$ sources per pointing, a number density of $\sim 0.5\,$arcmin$^{-2}$. We run the PyBDSF source finder on these images and for the resultant catalogue run the SuperCALS pipeline. Given that the \sersic index is fixed to $n=1$ for the input sources, we only run the \imshape parts of SuperCALS in this mode, in contrast to the real data where the \sersic index is chosen from the best fitting one across multiple \imshape runs.

\begin{figure*}
\includegraphics[width=0.5\textwidth]{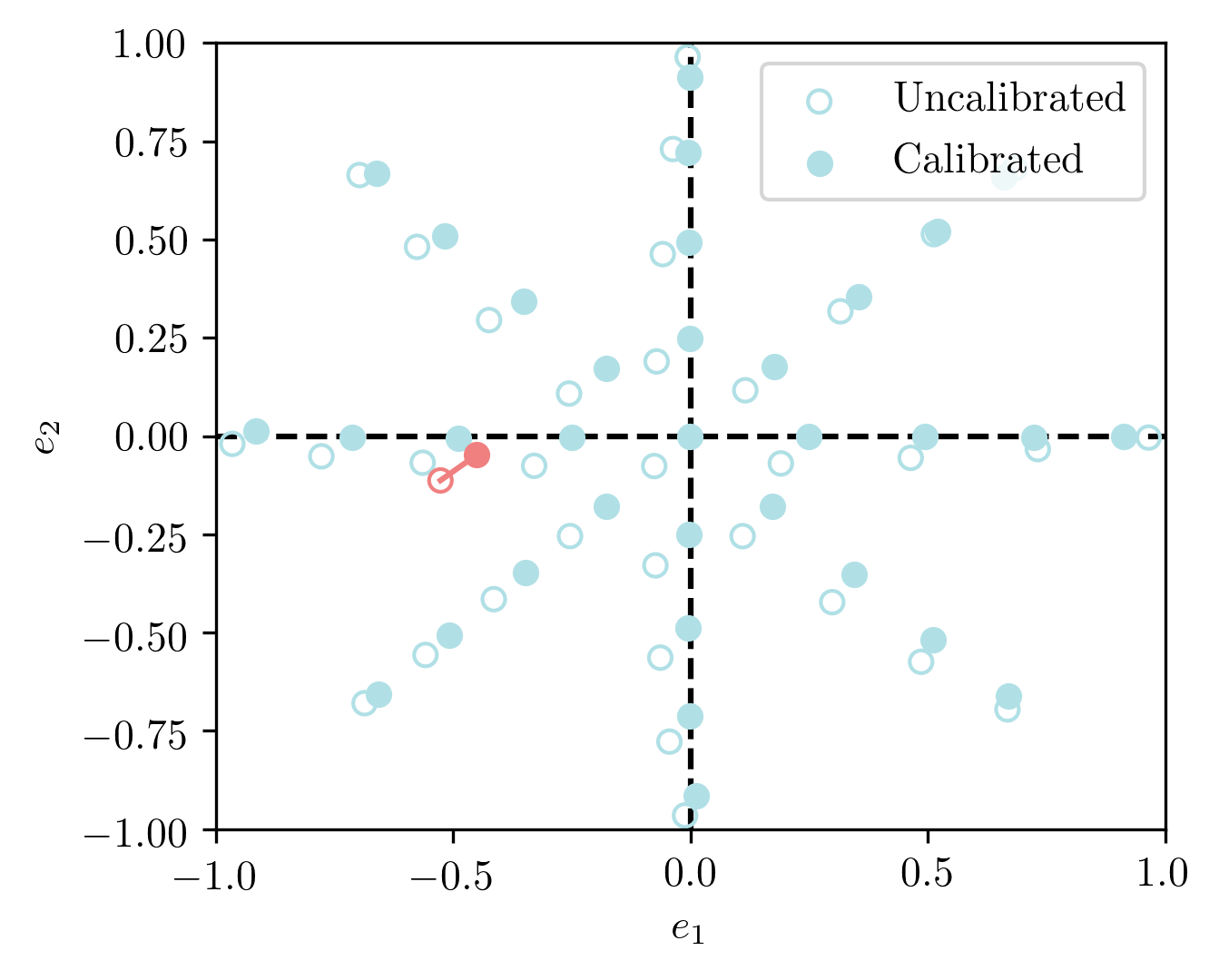}\includegraphics[width=0.5\textwidth]{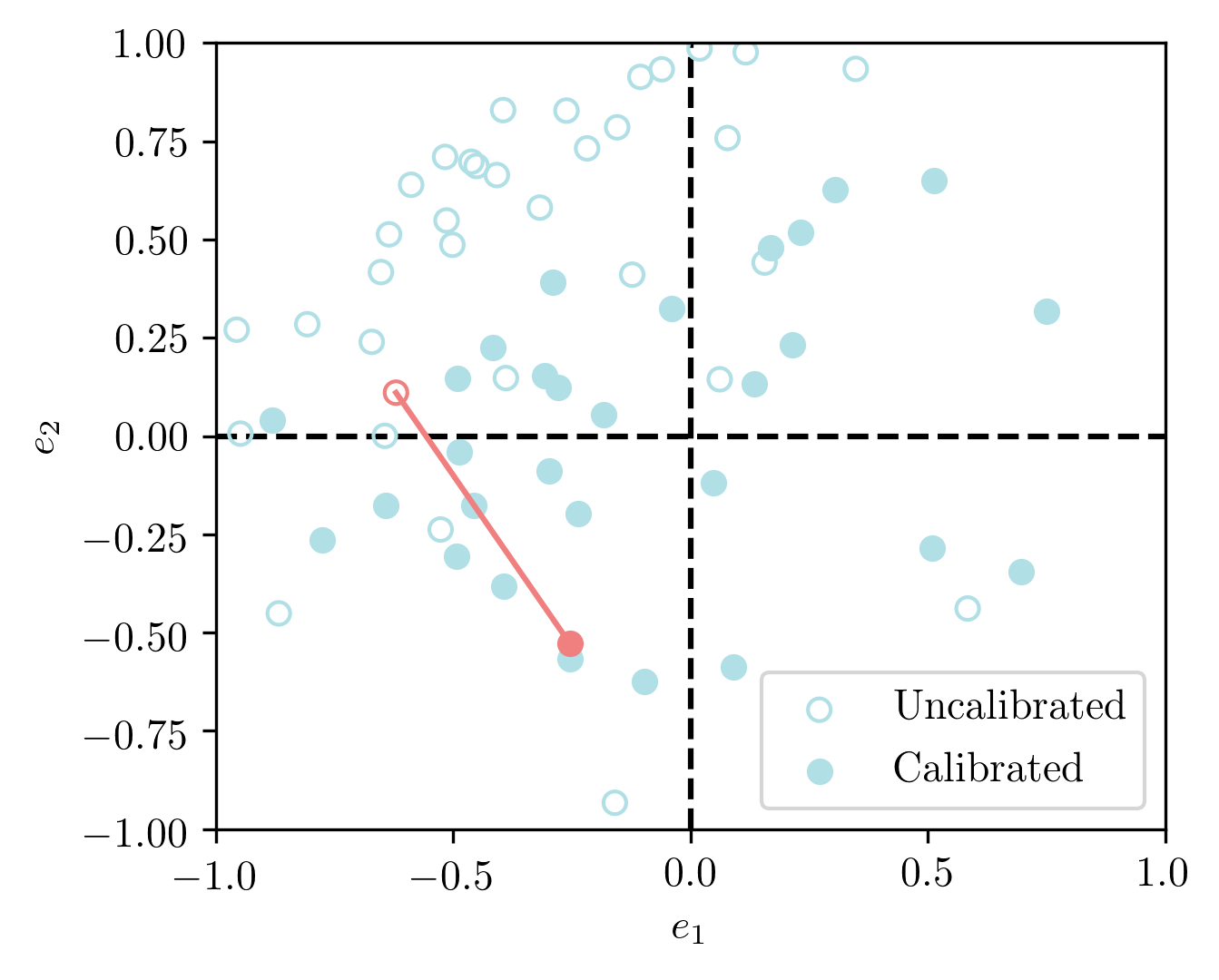}
\caption{`Calibration crosses' from the SuperCALS method. Unfilled blue circles show the recovered ellipticies of simulated sources injected into the \clean residual image at the location of the real source. Filled blue circles show the calibrated ellipticity measurements for these sources. Red unfilled and filled points show the ellipticity of the real source, before and after calibration. \emph{Left} shows a source for which calibration is considered a success, \emph{right} a source for which calibration is considered a failure.}
\label{fig:supercals-cross}
\end{figure*}

In \cref{fig:supercals-cross} we illustrate the calibration of an individual source ellipticity by the SuperCALS process. Unfilled blue points show the measured ellipticities of the sources injected into the \clean residual image. The difference between the injected and recovered values of ellipticity are used to calculate the calibration surface using \cref{eqn:supercals_bias_surface}, with filled circles showing the calibrated positions of the injected sources after the bias correction is applied (note that input positions are not perfectly recovered due to the interpolation). As can be seen in the left panel, the significant biases represented by the distorted shape of the eight-pointed cross can be corrected to a high level. We also show in red the uncalibrated (unfilled) and calibrated (filled) location of the observed source's ellipticity.

This procedure works for the vast majority of sources, but a number of sources with particularly significant image plane noise artefacts have `calibration crosses' (the data points in \cref{fig:supercals-cross}) which are highly distorted, to the point where they no longer form an obvious cross (i.e. as in the right panel). We find in general these sources cannot then be succesfully calibrated, with crosses remaining significantly distorted after calibration, and hence we exclude them from our shape catalogue. This is done by measuring the total square distance $d_{\times}$ between the uncalibrated injection points in the calibration cross (unfilled blue circles in \cref{fig:supercals-cross}) and the input ellipticity values for the injected sources. For sources which are clear visual failures, with the cross shape not being recovered, this value is in the range $\sim5-50$ (with numbers decreasing sharply as $d_{\times}$ increases), and we remove all sources with $d_{\times}>5$, which corresponds to a fraction of $\approx8$\%.

\begin{figure}
\includegraphics[width=0.5\textwidth]{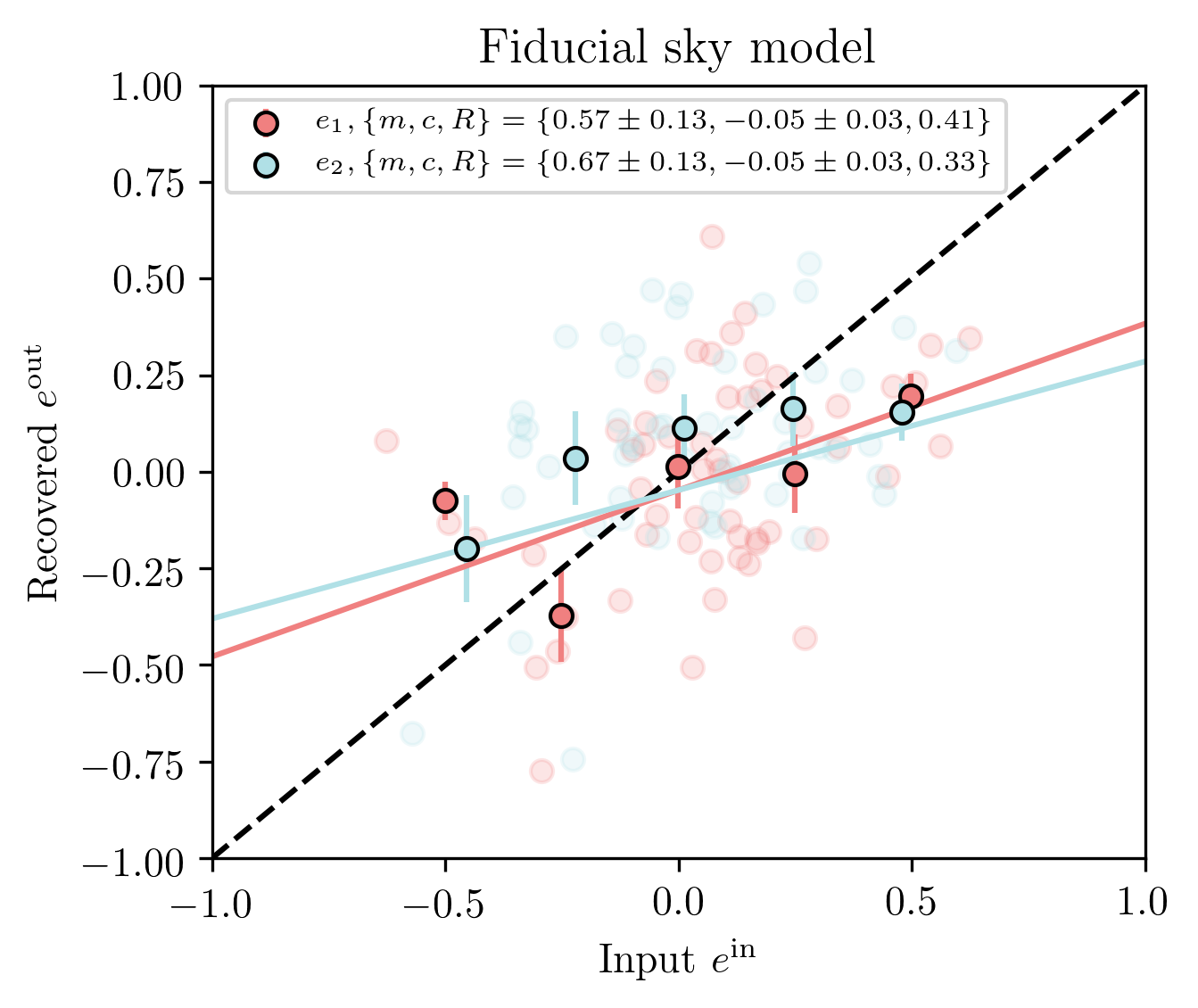}
\caption{Input vs recovered source ellipticity for the simulated J28 pointing of the VLA DR1 data using SuperCALS, with the fiducial T-RECS sky model. The legend indicates the intercept and slope of the best-fitting linear relation, and the estimated correlation coefficient between the $e^{\rm in}$ and $e^{\rm out}$ values.} 
\label{fig:supercals-1times1}
\end{figure}

\begin{figure}
\includegraphics[width=0.5\textwidth]{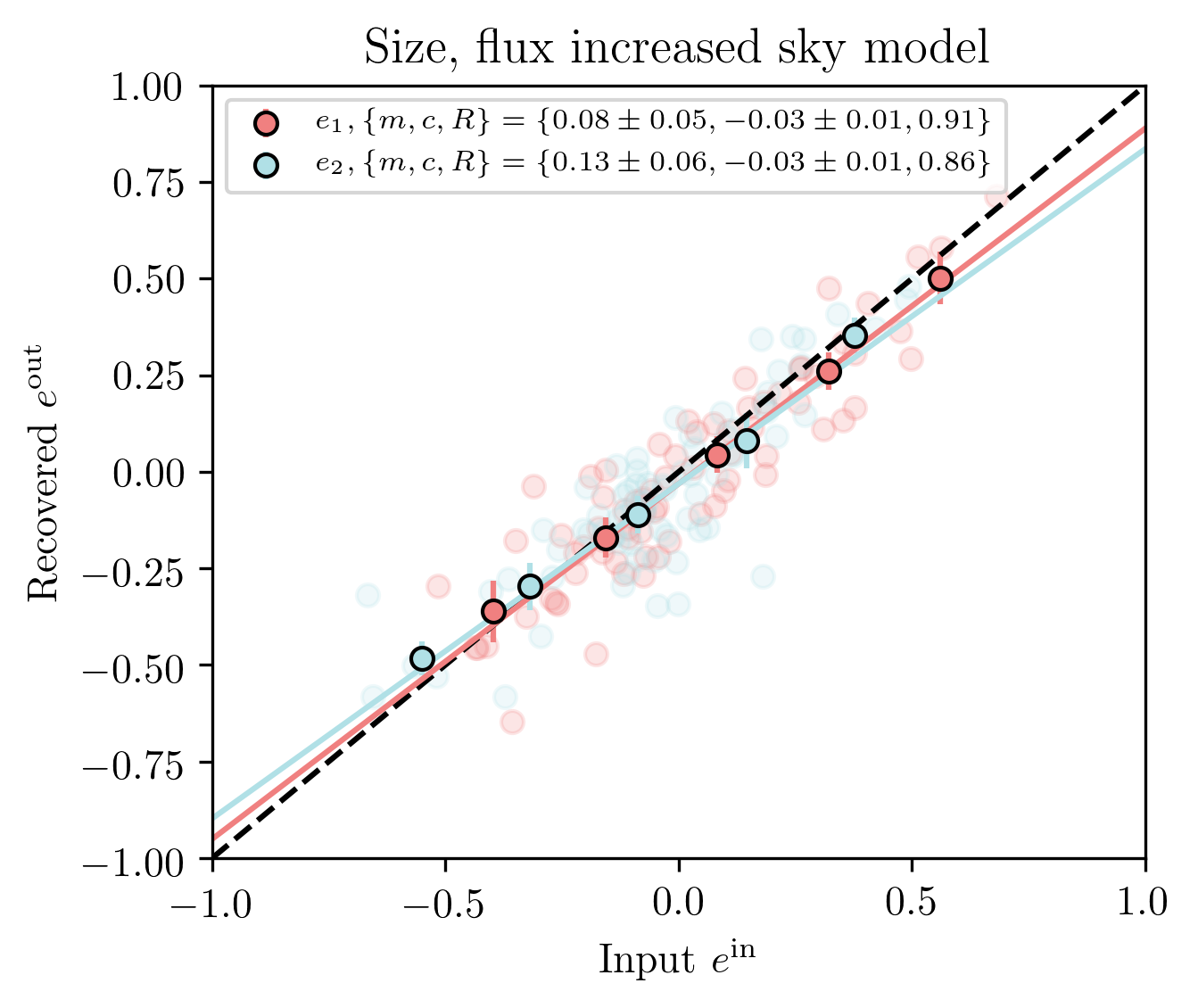}
\caption{Input vs recovered source ellipticity for the simulated J28 pointing of the VLA DR1 data using SuperCALS, with the fiducial T-RECS sky model modified so that each source is three times larger and 100 times brighter, demonstrating the ability of the method to recover unbiased source morphology. The legend is as in \cref{fig:supercals-1times1}.}
\label{fig:supercals-3times100}
\end{figure}

We have applied this method to a number of simulations, varying the input properties of the input T-RECS sky model. For the fiducial sky model specified by T-RECS, the shape measurement recovery (the input ellipticities plotted against the calibrated measured ellipticities) for a single pointing region of the VLA data is presented in \cref{fig:supercals-1times1}, and for a sky model in which we increase both sizes (by a factor of three) and fluxes (by a factor of 100) in \cref{fig:supercals-3times100}. \ih{For these sources, the SNR defined as the ratio of the peak recovered source flux to the RMS noise is $12 < {\rm SNR}_{\rm peak} < 580$ and for the SNR defined as the ratio with the total recovered source flux is $80 < {\rm SNR}_{\rm total} < 1200$.} For the VLA data, which has a relatively large restoring beam ($1.9 \times 1.5\,$arcsec) compared to the typical size ($\sim1\,$arcsec) of T-RECS star-forming galaxies, the shapes of the fiducial sky model are not recovered well. This is in contrast to the case of the bright, large source sky model, which demonstrates the ability of the SuperCALS method to recover unbiased shapes.

We quantify this with the linear ellipticity bias model $e^{\rm obs} - e^{\rm inp} = me^{\rm inp} + c$ for the recovered shapes, and the Pearson correlation coefficient $R$ between the input and output ellipticities, the measured values for which are shown in the figure legends. For the fiducial sky model, the correlation coefficient is low (consistent with zero for the 59 data points here if they are assumed uncorrelated) and the SuperCALS method is therefore unable to significantly measure the input shapes in this case (i.e. there is little morphology information preserved in the image for it to measure). When reporting measured shapes from SuperCALS for the radio sources in \cref{sec:radio_meas-shapes} we also apply the linear bias correction derived from \cref{fig:supercals-1times1} as a \emph{population-level} calibration in addition to the \emph{source-level} calibration part of SuperCALS and described above.

As an additional shape measurement method to SuperCALS, we also provide the initial, uncorrected \imshape-only shape measurement of the VLA source shapes.

\subsection{Radio shape analysis}
\label{sec:radio_meas-shapes}
\begin{figure}
\includegraphics[width=0.475\textwidth]{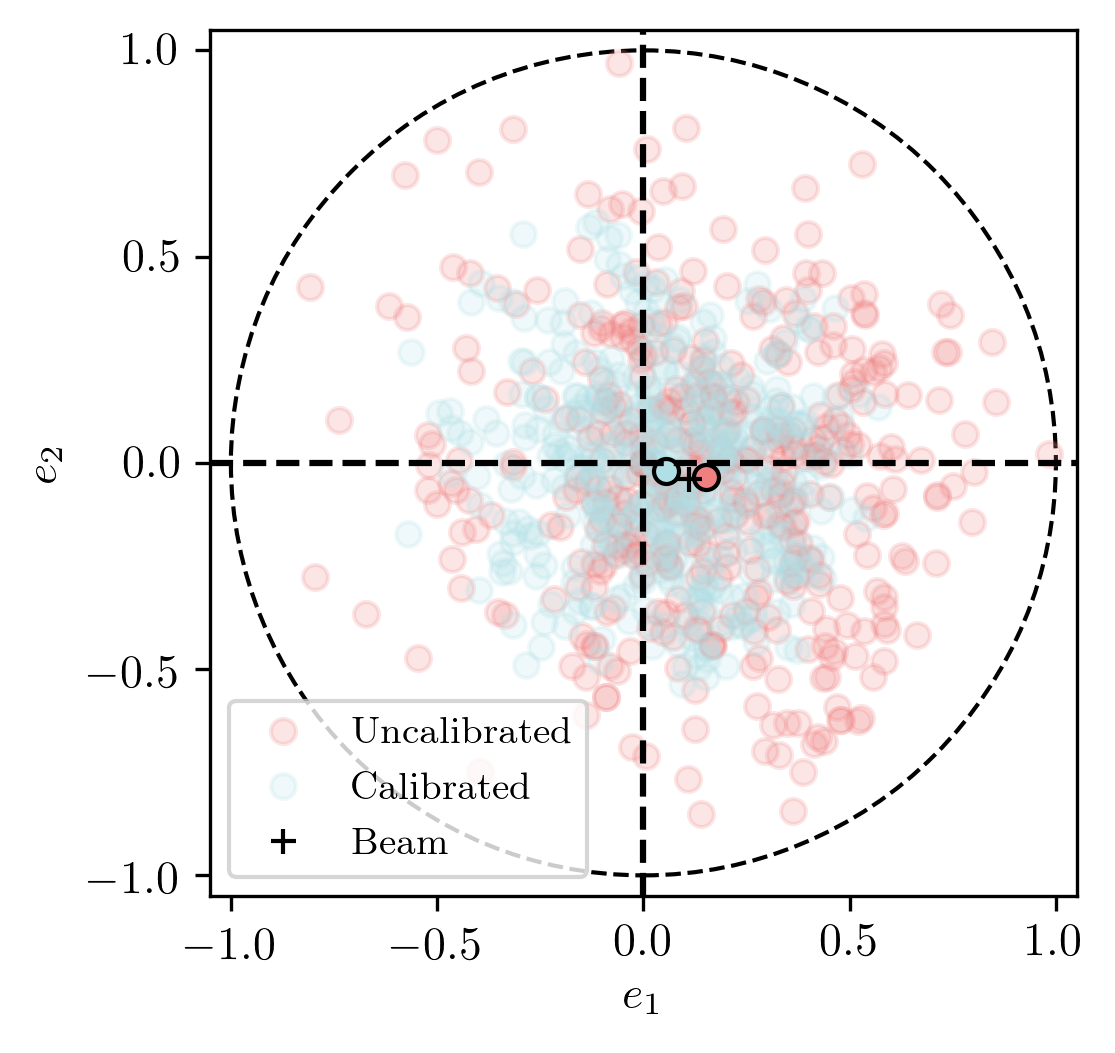}
\caption{Measured and calibrated ellipticities from the SuperCALS shape measurement method for the weak lensing radio catalogue from the VLA data. Dashed lines represent the unit circle in which physical ellipticities lie, and the points with solid black outlines are the averages for uncalibrated and calibrated ellipticities.}
\label{fig:supercals-dr1-e1e2}
\end{figure}

\begin{figure*}
\includegraphics[width=0.95\textwidth]{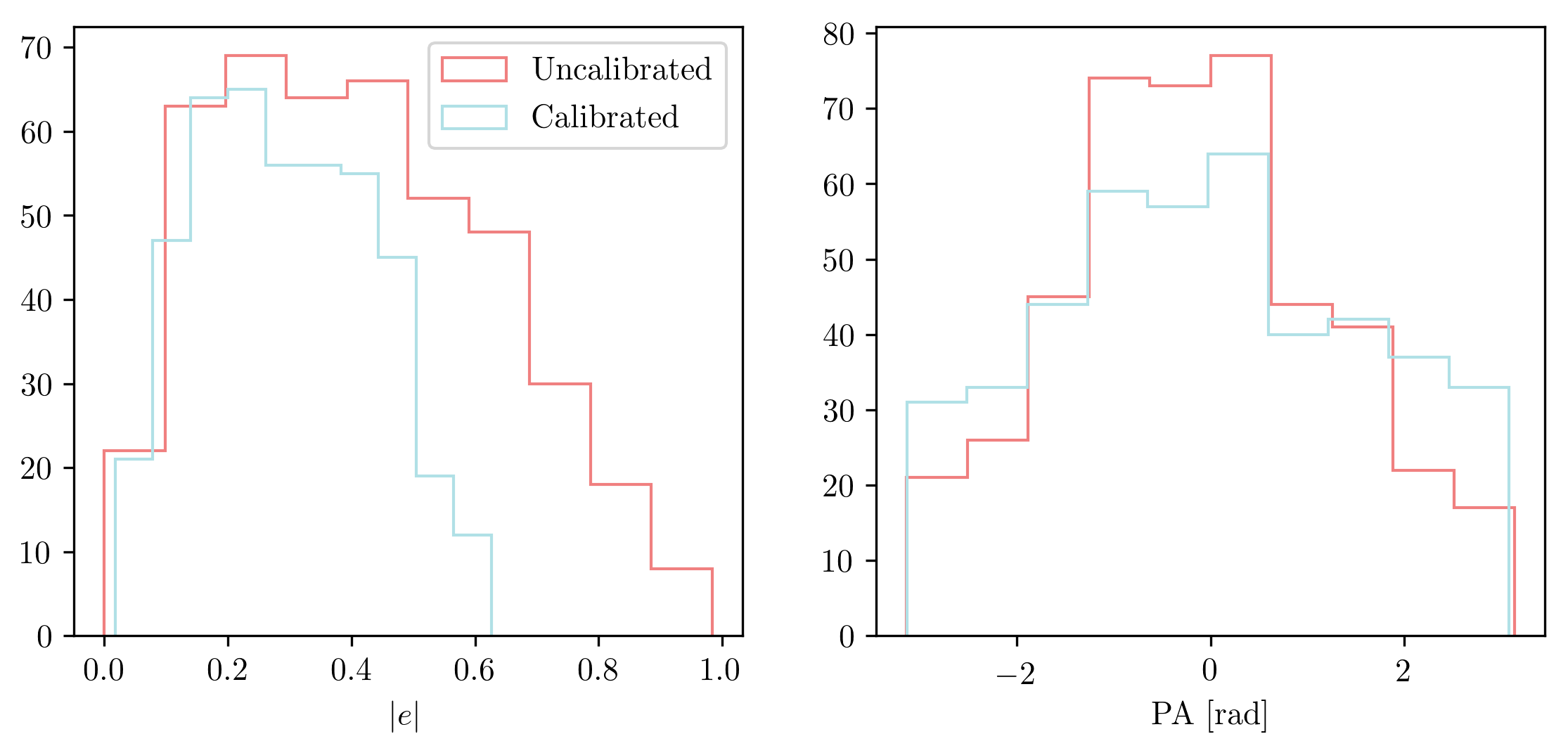}
\caption{Ellipticity modulus and position angle histograms from the SuperCALS shape measurement method for the weak lensing radio catalogue from the VLA data.}
\label{fig:supercals-dr1-mode_pa}
\end{figure*}

\begin{figure}
\includegraphics[width=0.5\textwidth]{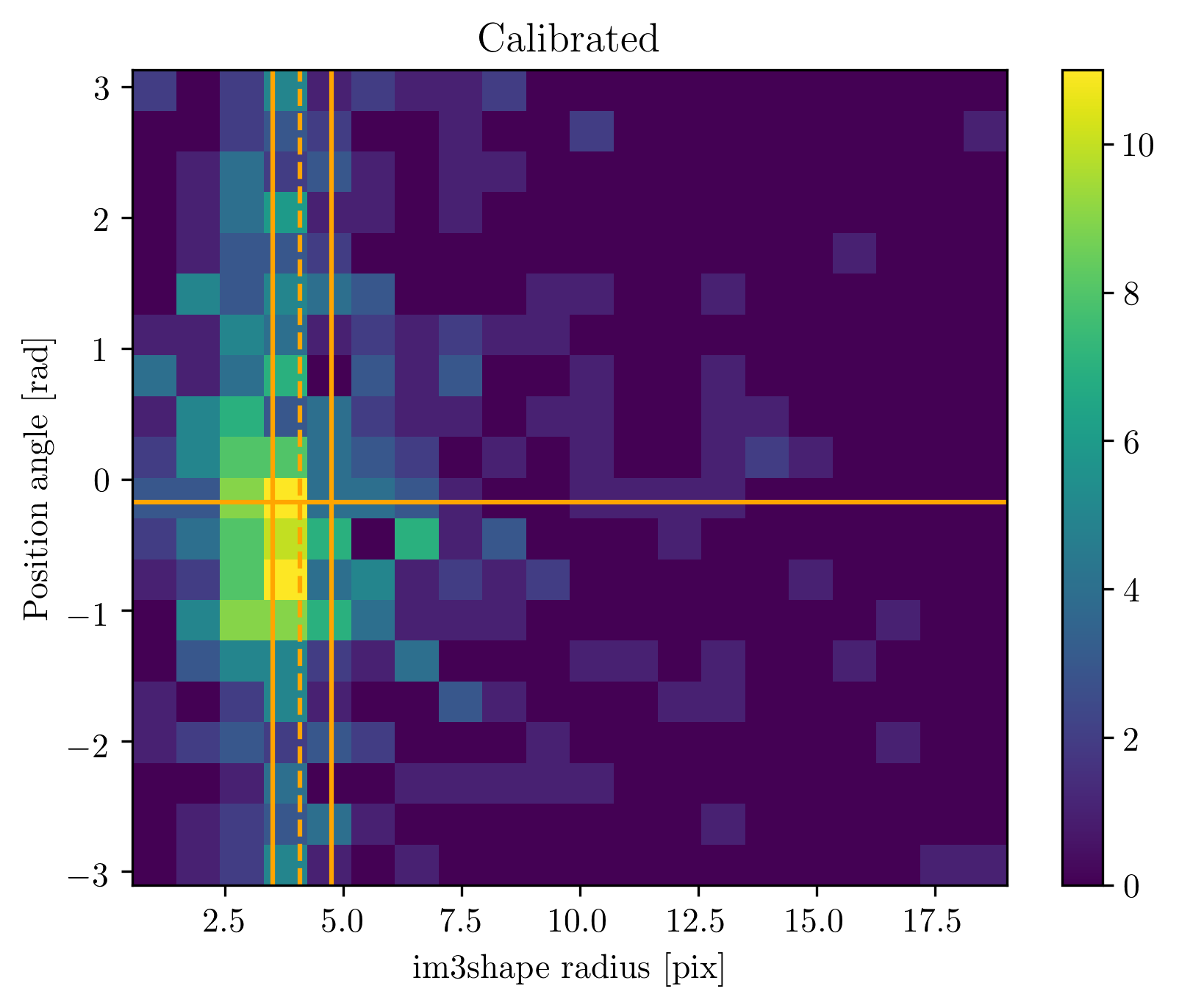}
\caption{\imshape radii compared to the restoring beam position angle for the VLA data. Orange lines show the position angle of the restoring beam and its size (minor axis, major axis, and the geometric mean of these as the dashed line). This demonstrates that the size and shape of the beam is still dominating the morphology recovered for these sources.}
\label{fig:im3shape_radii}
\end{figure}

We apply the SuperCALS shape measurement described above to our VLA-only weak lensing catalogue, described in \cref{sec:radio_meas-cataloguep-VLA}. We include in our catalogue both calibrated and uncalibrated measurements of the galaxy ellipticities $e_1$ and $e_2$ (the uncalibrated measurement corresponds to simply running \imshape on the image, the same method as used in \cref{sec:optical_meas-shapes} for the optical data). These shape measurements are plotted in \cref{fig:supercals-dr1-e1e2} and histograms of their ellipticity modulus $|e| = \sqrt{e_1^2 + e_2^2}$ (for which we expect a Rayleigh-like distribution) and position angle $\mathrm{PA} = 0.5\tan^{-1}(e_2/e_1)$ (for which we expect a flat uniform distribution) in \cref{fig:supercals-dr1-mode_pa}.
Inspection of the uncalibrated shape distributions in \cref{fig:supercals-dr1-mode_pa} leads us to doubt that credible conclusions can be drawn on morphological information from this data set. The histogram of position angle shows a significant peak at the position angle of the beam, in spite of the resolution cut imposed in \cref{sec:radio_meas-cataloguep-VLA}. Upon inspection of the \imshape measured source sizes, we find many are consistent with being smaller than the restoring beam used for the VLA imaging, as shown in \cref{fig:im3shape_radii}, which again demonstrates the significant peak at the beam position and size.
This strongly indicates that the morphology of many of the shapes in this image are still dominated by the restoring beam imposed in the imaging step. To illustrate this further, in \cref{fig:im3shape_radii} we show a 2D histogram of the measured PSF-deconvolved source position angle and radius, this time measured in pixels by \imshape while performing the shape fitting. Orange lines represent the shape of the restoring beam used in the VLA imaging process. As can be seen, even though these sources pass the resolution cut from \textsc{PyBDSF} measurements in \cref{sec:radio_meas-cataloguep-VLA}, they do not appear resolved when measured with \imshape and are still dominated by the beam shape. For further discussion of the ability of our data to constrain the size and \sersic profile of these sources see Section 5.2 of Paper I. These sources may be from unresolved sources interacting with noise peaks in the image, causing their size to be artificially `upscattered', an effect expected in radio interferometer images \citep[see e.g.][]{2019ApJ...883..204T}.

Because of this residual effect of the beam position angle on the shape, we apply a further correction to the source ellipticities. A linearly biased galaxy ellipticity measurement can be rotated into the frame of a PSF with known position angle $\alpha_{\rm PSF}$:
\begin{equation}
    \exp(-i2\alpha_{\rm PSF})e_{\rm obs} = (1 + m)\exp(-i2\alpha_{\rm PSF})e_{\rm true} + c,
\end{equation}
and, if $m$ and $c$ can be reliably estimated then we can correct for the known PSF ellipticity:
\begin{equation}
    e_{\rm corrected} = \frac{e_{\rm obs} - \exp(i2\alpha_{\rm PSF})c}{1 + m}.
\end{equation}
We therefore use the values for $m$ and $c$ derived from our simulations in \ih{\cref{sec:radio_meas-supercals-sims}} along with the known restoring beam position angle of $80\,\deg$ (imposed as part of the VLA imaging process). Shapes presented (as `Calibrated') in \cref{fig:supercals-dr1-e1e2} and \cref{fig:supercals-dr1-mode_pa} and in the shape catalogue used in \cref{sec:power_spectra} have this correction applied.

For the analysis presented here, weak lensing measurements are still dominated by shot noise due to the low number density of sources, dominating over even this significant systematic. In the next section we proceed to measure the primary weak lensing observable of the shear power spectrum inferred from these shapes. For the full data release with higher number density of galaxies and hence lower shot noise, we expect that improvements in the imaging procedure may remove this systematic feature. The addition of smaller spatial scales from the \emerlin data will also raise the level of morphological information in the image.

\section{Shear power spectra}
\label{sec:power_spectra}
From the galaxy shapes measured in \cref{sec:optical_meas,sec:radio_meas} we measure the angular shear power spectra, $C_\ell^{\gamma(i)\,\gamma(j)}$. The two-point statistics of the shear field are sensitive to cosmological parameters through the underlying matter overdensity power spectrum $P_\delta(k)$, which is a linear function of the Gaussian primordial perturbations on large scales. For the shear power spectrum observable used here \citep[e.g.][]{Bartelmann2001}:
\begin{equation}
  C_\ell^{\gamma(i)\,\gamma(j)} = \frac{9 H_0^4 \Omega_{\mathrm{m}}^2}{4 c^4} \int_{0}^{\chi_{\mathrm{h}}} d\chi \frac{g^i(\chi)\,g^j(\chi)}{a^2(\chi)} P_\delta \left(\frac{\ell}{f_K (\chi)}, \chi \right),
    \label{eq:powspecfull}
\end{equation}
where $H_0$ is the Hubble constant, $\Omega_\mathrm{m}$ is the total matter density, $c$ is the speed of light in a vacuum, $\chi$ is the comoving distance, $a (\chi)$ is the scale factor of the Universe and $f_K (\chi)$ is the comoving distance ($f_K (\chi) = \chi$ for a flat Universe). The kernels, $g^{i, j} (\chi)$ describe the relative contributions of the two galaxy samples to the lensing signal, and are given by
\begin{equation}
  g^{i} (\chi) = \int_{\chi}^{\chi_{\mathrm{h}}} d\chi^\prime n_{i} (\chi^\prime) \frac{f_K (\chi^\prime - \chi)}{f_K (\chi^\prime)},
    \label{eq:lensing kernels}
\end{equation}
where $n_i (\chi)$ is the distribution of galaxies, as a function of comoving distance, for galaxy sample $i$, and the integral extends to the horizon, $\chi_{\mathrm{h}}$. Here the sample labels $i,j$ take on values `O' for the optical sample and `R' for the radio sample, giving the three power spectra $C_\ell^{\rm RR}, C_\ell^{\rm OO}$ and $C_\ell^{\rm RO}$.

In order to calculate the power spectra we use the publicly available\footnote{\url{https://bitbucket.org/fkoehlin/qe\_public}} flat-sky maximum likelihood power spectrum estimation code fully described in \cite{2016MNRAS.456.1508K, 2017MNRAS.471.4412K}. This code is a development of the algorithm presented in \cite{2001ApJ...554...67H} to allow for a tomographic analysis between several pairs of redshift bins. We do not apply a tomographic shear analysis in this work. Instead, we perform a 2D cosmic shear analysis between two shear maps, where all background sources reside in a single, very broad, redshift bin.

When estimating the power spectrum \cref{eq:powspecfull} from a finite sample of galaxies the shear, $\gamma$ is estimated from the ensemble average of the measured ellipticities of a number of sources $N$ within a small region or `shear pixel' of the sky:
\begin{equation}
  \widehat{\gamma} = \frac{1}{N} \sum^{N}_i \epsilon_i.
  \label{eq:avg_shear}
\end{equation}
We pixelate our shear maps with a side length of $\theta_{\rm pix} = 3\,{\rm arcmin}$. The radio and optical shear catalogues have galaxy number densities of $n_{\rm gal}^{\rm R} \approx 0.5\:\mathrm{arcmin}^{-2}$ and $n_{\rm gal}^{\rm O} \approx 19\:\mathrm{arcmin}^{-2}$, leaving each shear pixel with an average of $\approx153$ optical sources and $\approx4.5$ radio sources.

The estimate of shear from averaging down galaxy shapes with $\langle\epsilon\rangle = 0$ but $\langle\epsilon^2\rangle \neq 0$ on a given angular scale $\ell$ then has a shot noise term, related to the number density of available galaxy shapes, ${n_{\mathrm{gal}}^i}\:\mathrm{arcmin}^{-2}$ and the expected covariance of the intrinsic (i.e. before lensing) galaxy shapes:
\begin{align}
  \mathcal{N}^{i j} &= \frac{1}{n_{\rm gal}^{i}n_{\rm gal}^{j}}\langle \sum_{\alpha \in i}\epsilon_\alpha  \sum_{\beta \in j}\epsilon_\beta \rangle \nonumber \\
&= \frac{n_{\rm gal}^{i j}}{n_{\rm gal}^{i}n_{\rm gal}^{j}}\mathrm{cov}(\epsilon_{i}, \epsilon_{j}).
  \label{eq:cov_shear}
\end{align}
where $n_{\rm gal}^{i j}$ is the number of galaxies common to both samples. Here we assume negligible overlap between the radio and optical galaxy samples, meaning the cross-noise power spectrum is also negligible, $\mathcal{N}^{\rm RO} = 0$ \citep[which will also be a good approximation on the angular scales considered here, as discussed in][]{2019MNRAS.488.5420H}. However, for the auto power spectra where $i=j$, \cref{eq:cov_shear} gives $\mathcal{N}^{\rm OO} = \sigma^{2}_{\epsilon_{\rm O}}/n_{\rm gal}^{\rm O}$ and $\mathcal{N}^{\rm RR} = \sigma^{2}_{\epsilon_{\rm R}}/n_{\rm gal}^{\rm R}$, where $\sigma_{\epsilon_{i}}$ is the dispersion of intrinsic galaxy ellipticites in the $i^{\rm th}$ sample.

\subsection{Band power selection}
\label{sec:power_spectra-bandpower_selection}
The nominal multipole ranges for power spectra extraction were selected following the prescription outlined in \cite{2017MNRAS.471.4412K}. While the method of extracting multipoles is the same for both the DR1 and full regions, we present spectra for both because of the differing areas.

The largest multipole available to be extracted from the shear maps is set by the shear pixel side length, $\theta_{\rm pix} = 3\,{\rm arcmin}$, corresponding to a multipole of $\ell_{\rm pix} = 7200$. The smallest multipole available is determined by the survey areas. The DR1 region covers an area of $0.26\,\degsq$, so we choose a survey side length of $\theta_{\rm max}^{\rm DR1} = \sqrt{0.26}\,\deg$, corresponding to $\ell_{\rm field}^{\rm DR1} = 710$. We have also run the power spectrum estimator on simulations, see \cref{sec:power_spectra-simulations}, which cover an area of $1\,\degsq$. For these runs, the survey side length is $\theta_{\rm max}^{1\,\degsq} = 1\,\deg$, corresponding to $\ell_{\rm field}^{1\,\degsq} = 360$. To account for DC offset effects and/or ambiguous modes (modes which cannot be distinguished into $E$- and $B$-modes), we also include ``junk" multipole bins at lower multipoles with a lowest multipole of $\ell_{\rm min} = 100$. Since this bin contains unreliable band power measurements, we discard it in any further analysis.

The widths of intermediate multipole bins were set to at least $2 \ell_{\rm field}$ to minimise correlations between the band powers \citep{2001ApJ...554...67H}. These widths corresponded to 1420 and 720 for the $0.26\,\degsq$ DR1 and $1\,\degsq$ binnings, respectively. The maximum multipole of the highest-$\ell$ band power for both binnings was extended to $2 \ell_{\rm pix} \approx 14,400$, to absorb any effects resulting from the highly oscillatory behaviour of the pixel window function on scales close to and larger than $\ell_{\rm pix}$ \citep{2017MNRAS.471.4412K}. This highest-$\ell$ band power was also labelled a ``junk" bin and was discarded for further analysis. Moreover, we have also followed \cite{2017MNRAS.471.4412K} in discarding the second-to-last bin when interpreting the results.

\Cref{tab:bandpowerselection} lists the resulting band power definitions. Only bins 2, 3 (and 4 for the $1\,\degsq$ binning) contain reliable band powers for the reasons mentioned above. The table also contains approximate real-space $\theta$-ranges for each bin. These ranges only serve as an approximation of the real-space scales probed by each band power, and should not be used to directly compare the power spectra measurements to real-space correlation function analyses for reasons that are discussed in \cite{2017MNRAS.471.4412K}.

\begin{table}
	\centering
	\caption{Band power intervals used for the power spectra extraction with the two different binnings corresponding to the $0.26\,\degsq$ DR1 region and the $1\,\degsq$ simulations. Only bins 2, 3 (and bin 4 for the $1\,\degsq$ binning) were retained for further analysis, see \cref{sec:power_spectra-bandpower_selection}. The extraction of the full $1.53\,\degsq$ optical area used the $1\,\degsq$ binning, see \cref{sec:power_spectra-optical_full}. Bands shown in grey are measured but not used in the analysis.}
	\label{tab:bandpowerselection}
	\begin{tabular}{ccccc}
		\hline
		Band & \multicolumn{2}{c}{$0.26\,\degsq$ binning} & \multicolumn{2}{c}{$1\,\degsq$ binning}\\
        No. & $\ell$-range & $\theta$-range & $\ell$-range & $\theta$-range \\
         & & [arcmin] & & [arcmin] \\
		\hline
		\color{mygrey}{1} & \color{mygrey}{100-710} & \color{mygrey}{216.0-30.4} & \color{mygrey}{100-360} & \color{mygrey}{216.0-60.0}\\
		2 & 711-2499 & 30.4-8.6 & 361-1099 & 59.8-19.7\\
		3 & 2500-4999 & 8.6-4.3 & 1100-2499 & 19.6-8.6\\
        4 & \color{mygrey}{5000-7200} & \color{mygrey}{4.3-3.0} & 2500-4499 & 8.6-4.8\\
        \color{mygrey}{5} & \color{mygrey}{7201-14400} & \color{mygrey}{3.0-1.5} & \color{mygrey}{4500-7200} & \color{mygrey}{4.8-3.0}\\
        \color{mygrey}{6} & \color{mygrey}{-} & \color{mygrey}{-} & \color{mygrey}{7201-14400} & \color{mygrey}{3.0-1.5}\\
		\hline
	\end{tabular}
\end{table}

\subsection{Simulations}
\label{sec:power_spectra-simulations}
To estimate the uncertainties in the band power measurements, we used the simulated $1\,\degsq$ \ih{supercluster} shear maps introduced in \cref{sec:radio_meas-simuclass-shear}. These simulations were also used to extract the expectation band powers for a cluster field and to assess the performance of the power spectrum estimator in the presence of the masking and spatial distribution of sources found in the real data.

In total, 61 \ih{supercluster} sub-volume simulations were used, each covering an area of $1\,\degsq$ and split into 20 redshift bins ranging from $z_{\rm min}=0.1$ to $z_{\rm max}=2.0$ in increments of $\Delta z = 0.1$. The shear maps were made with a resolution of 0.1 arcmin pixel$^{-1}$ and dimensions of 600$\times$600 pixels.

\begin{figure}
\includegraphics[width=0.5\textwidth]{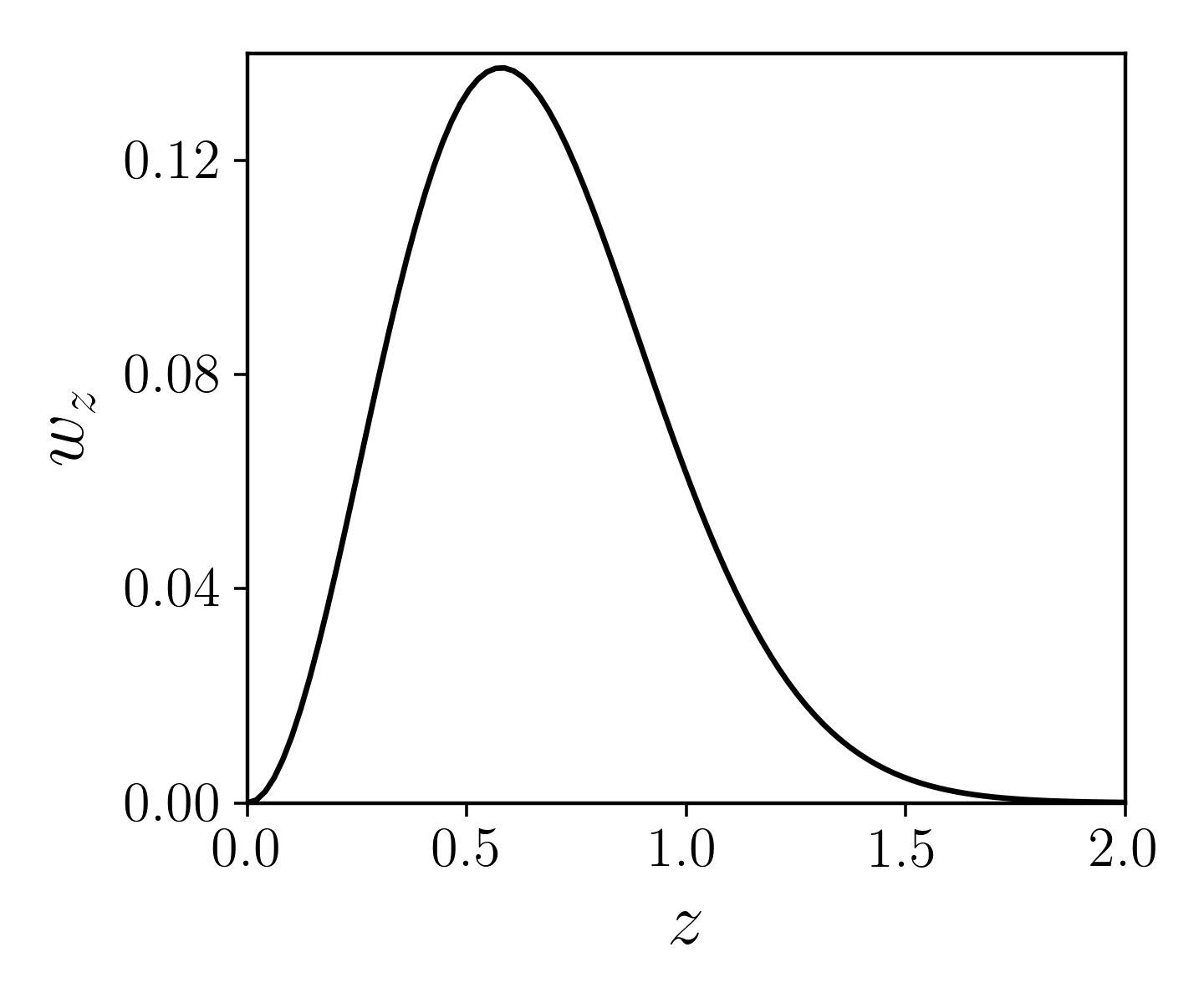}
\caption{The redshift distribution used to weight the simulated shear maps, given by equation~(\ref{eq:lima_redshift_4}) with values of $z_d(i) = -0.1219$ and $\sigma_d(i) = 0.636$}
\label{fig:redshift_dist_weights}
\end{figure}

We combine each set of 20 redshift slices by weighting them by redshift according to
\begin{equation}
  \overline{\gamma_{1,2}(z)} = \sum_{z=0.1}^{z=2.0} w_z \gamma_{1,2}(z),
  \label{eq:shear_map_weighting}
\end{equation}
where the horizontal bar indicates the weighted average shear for each $0.1\times0.1$ arcmin$^2$ pixel, applied on a component-by-component basis. The sum is over the redshift range of the input shear maps and the weights are described by \citep{2008MNRAS.390..118L}:
\begin{equation}
  w_z = P( z | i^\prime ) = B z^2 \exp{\left \{ -\left [ \frac{z-z_d(i^\prime)}{\sigma_d(i^\prime)} \right ]^2 \right \} },
  \label{eq:lima_redshift_4}
\end{equation}
for a given median $i^\prime$-band magnitude, $i^\prime$. The normalisation factor, $B$ is obtained by requiring that
\begin{equation}
  \sum_{z_{\rm min} = 0.1}^{z_{\rm max} = 2.0} P( z | i^\prime ) = 1
  \label{eq:redshift_normalise}.
\end{equation}
The constants, $z_d(i^\prime)$ and $\sigma_d(i^\prime)$ were estimated by using a combination of our measured median $i^\prime$-band magnitude of $i^\prime=22.57$ and values given in \cite{2008MNRAS.390..118L}, which made use of mock DES catalogues. This gave values of $z_d(i^\prime) = -0.1219$ and $\sigma_d(i^\prime) = 0.636$ and the resulting redshift distribution of $w_z$ is shown in \cref{fig:redshift_dist_weights}.

This process created 61 redshift-weighted shear maps, which were used as the basis for both our optical and radio simulations. For the $0.26\, \degsq$ DR1 area with overlapping optical and radio regions, the band power uncertainties were estimated by sampling the simulated shear fields at the positions of real sources as listed in the optical and radio shape catalogues. For the $1.53\, \degsq$ full area, we use random positions scattered over $1\, \degsq$, with the same source number density as the full optical catalogue of $n_{\rm gal}^{\rm O} = 19\:\mathrm{arcmin}^{-2}$, and then scale the error bars accordingly. We choose random positions here since the full $1.53\, \degsq$ optical catalogue does not contain a continuous square region covering $1\, \degsq$. Details of the error bar scaling are given in \cref{sec:power_spectra-optical_full}.

Shape noise was added by randomly selecting real measured ellipticities from the shape catalogues, and adding them to the shear entries for the galaxies. This naturally replicates the measurement error contributions from the real catalogues, and in the case where real data positions are used, the spatial distributions of sources in the real data are incorporated, including for example the gaps in the optical data coverage due to masking.

Each set of 2$\times$61 mock data maps (optical and radio) was run through the power spectrum estimator using the same settings as the data runs, including the two different multipole bin sets described in \cref{sec:power_spectra-bandpower_selection}. Error bars were extracted from the standard deviation of the 61 recovered band powers in each multipole bin, where we assume a symmetrical distribution (not necessarily Gaussian) about each mean band power.

We also generate expectation band powers, $P_b^{\rm th}$ from the 61 redshift-weighted, noiseless shear maps. Ordinarily, one would convolve a theoretical cosmic shear signal, $C^{\rm Cluster}_\ell$ with the band power window functions, $W_{b\ell}$ \citep[e.g.][]{1999PhRvD..60j3516K, 2012ApJ...761...15L}:
\begin{equation}
P_b^{\rm th}  = \sum_\ell \frac{W_{b\ell}}{\ell} \frac{\ell(\ell+1)}{2\pi} C^{\rm Cluster}_\ell
\label{eq:band_expectation}
\end{equation}
for each multipole bin, $b$. This standard approach was demonstrated in e.g. \cite{2019MNRAS.488.5420H}, where the theoretical cosmic shear signal chosen was a well-defined, flat, 6-parameter $\Lambda$CDM model. However, since the theoretical model for a \ih{supercluster} field, $C^{\rm Cluster}_\ell$ cannot be well defined, we use the redshift-weighted shear maps with no shape noise or masking to obtain the expectation band powers computationally. Each band power expectation value was extracted as the mean of those measured from the 61 noiseless shear maps. This was done for both sets of multipole bin definitions listed in Table~\ref{tab:bandpowerselection}, since different binnings produce different band power window functions.

\subsection{Detection significances and signal-to-noise ratios}
\label{sec:power_spectra-D_and_S}
For each of the power spectrum measurements, we calculate a detection significance, which quantifies the significance of a lensing signal being above a null signal, and a signal-to-noise ratio, accounting for both noise and cosmic variance.

Detection significances were calculated according to
\begin{equation}
D^\phi = \sqrt{\sum_b \left ( \frac{\hat{P}_b^\phi}{\sigma_b^\prime} \right )^2},
\label{eq:power_detection_significance}
\end{equation}
where $\phi \in \{EE, BB\}$, the index $b$ runs over the number of bands included, given in Table~\ref{tab:bandpowerselection}, $\hat{P}_b^\phi$ are the measured $E$- or $B$-mode band powers, and $\sigma_b^\prime$ are the uncertainties on the band powers excluding cosmic variance. These uncertainties were estimated using the standard deviations of the $B$-mode band powers\footnote{The cluster simulations contain zero input $B$-mode signal. Therefore the run-to-run scatters of the recovered $B$-mode band power estimates are free from cosmic variance.} recovered from the cluster simulations,
\begin{equation}
\sigma_b^\prime = \sqrt{\left < \left ( P_b^{BB} \right )^2 \right > - \left < P_b^{BB} \right >^2},
\label{eq:power_det_noise}
\end{equation}
where the angled brackets denote an average over the cluster simulations. Correspondingly, signal-to-noise ratios were calculated according to
\begin{equation}
S^\phi = \sqrt{\sum_b \left ( \frac{\hat{P}_b^\phi}{\sigma_b^\phi} \right )^2},
\label{eq:power_signal-to-noise}
\end{equation}
where $\sigma_b$ are now the uncertainties including both measurement noise and cosmic variance. The $\sigma_b$ values were estimated using the standard deviations of the $\phi$-mode band powers recovered from the cluster simulations:
\begin{equation}
\sigma_b^\phi = \sqrt{\left < \left ( P_b^\phi \right )^2 \right > - \left < P_b^\phi \right >^2}.
\label{eq:power_snr_noise}
\end{equation}
Note that equations~(\ref{eq:power_detection_significance}) and (\ref{eq:power_signal-to-noise}) are identical when calculating $B$-mode values.

\subsection{Optical power spectra for the full SuperCLASS region}
\label{sec:power_spectra-optical_full}
For the full $1.53\,\degsq$ SuperCLASS region, we apply a binning which is appropriate for the $1\,\degsq$ area. This conservative choice is motivated by the fact that the cluster simulations only cover an area of $1\,\degsq$, as discussed in \cref{sec:power_spectra-simulations}. Hence, only theoretical band powers and error bar estimates for a $1\,\degsq$ binning can be extracted. This does not mean that we lose any information on the scales that we do probe, since all the $1.53\,\degsq$ data is used. The conservative binning only means that we do not probe lower $\ell$-modes, which would extend down to $\ell_{\rm field}^{1.53 {\rm deg}^2} \approx 290$ for a $1.53\,\degsq$ binning scheme.

The difference in area between the simulations and the data also mean the error bars should be scaled. To first order, the error bars of a given survey scale according to \citep[e.g.][]{1992ApJ...388..272K}
\begin{equation}
\Delta \hat{C}_\ell = \sqrt{\frac{2}{(2\ell + 1) f_{\rm sky}}} \left ( \hat{C}_\ell + \frac{\sigma_\epsilon^2}{n_{\rm gal}} \right )
\label{eq:pow_spec_scaling}.
\end{equation}
Assuming the cosmological signal, $\hat{C}_\ell$ the shape noise variance, $\sigma_\epsilon^2$ and the galaxy number density, $n_{\rm gal}$ to be constant between the two areas, the ratio of error bars can be approximated as
\begin{equation}
\frac{\Delta \hat{C}_\ell^{1.53 \degsq}}{\Delta \hat{C}_\ell^{1 \degsq}} \approx \sqrt{\frac{f_{\rm sky}^{1 \degsq}}{f_{\rm sky}^{1.53 \degsq}}}
\label{eq:pow_spec_scaling_frac}.
\end{equation}
However, equation~(\ref{eq:pow_spec_scaling_frac}) does not account for the effect of differences in the masking between two maps. To model this effect, we create Gaussian random fields with the same 6-parameter, flat, $\Lambda$CDM input model used in~\cite{2019MNRAS.488.5420H}. These simulations were created over an area large enough to extract both a $1\,\degsq$ randomly sampled area and the full $1.53\,\degsq$ masked area. 100 sets of simulations were produced. The $1\,\degsq$ simulated shape catalogues were created in exactly the same way as for the cluster simulations: random positions with randomly-sampled real shape noise. The $1.53\,\degsq$ masked area simulations used positions from the real optical weak lensing catalogue, Fig.~\ref{fig:superclass_field}, and shape noise was included by randomly sampling shapes across the weak lensing catalogue. The differences in band power errors between the case where equation~(\ref{eq:pow_spec_scaling_frac}) is applied directly and when the simulations are used to also account for the mask are $\sim 10\%$ (with errors decreasing when the masking effect is included).

Each set of 2$\times$100 simulations was run separately through the power spectrum estimator and the standard deviations of the extracted band powers were measured. The ratios between the standard deviations of the two sets of band powers for each multipole bin were used to scale down the cluster simulation measured error bars.

\begin{figure}
\includegraphics[width=0.5\textwidth]{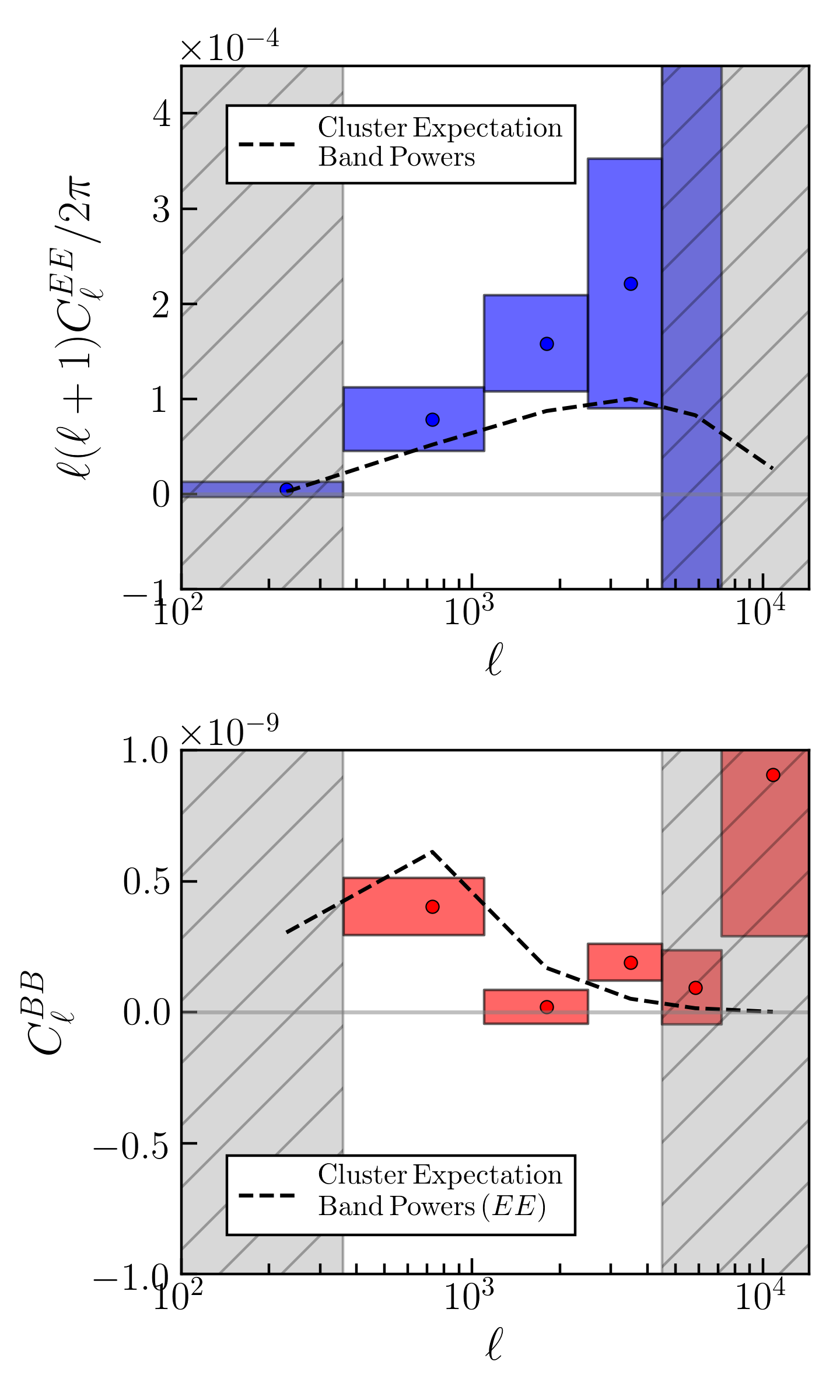}
\caption{Measured $E$- ({\it top}) and $B$-mode ({\it bottom}) optical-only power spectra for the full $1.53\,\degsq$ SuperCLASS region. The dashed curve illustrates the $E$-mode expectation band powers for a \ih{supercluster} region. Note the different vertical axis scales and factors of $\ell$ for the two panels. These power spectra measurements are quantified in Table~\ref{tab:power_spectra-DetSNR}.} 
\label{fig:full-optical-cl}
\end{figure}

In \cref{fig:full-optical-cl} we show the measured $E$- and $B$-mode power spectra for the full $1.53\,\degsq$ region covered by the optical data. Circles show the measured band powers for each range of $\ell$-modes, which are illustrated by the widths of the shaded regions. The vertical extents of the shaded regions represent the uncertainty in the estimated band powers. The band powers and uncertainties here correspond to a \ih{$9.9\sigma$} detection of a non-zero weak lensing $E$-mode power spectrum. We also measure a significant $B$-mode lensing signal at a detection significance of \ih{$4.6\sigma$}, indicating the presence of some residual $B$-modes. However, we note that these $B$-modes are measured at a smaller significance than the $E$-modes. These statements are quantified in Table~\ref{tab:power_spectra-DetSNR}.

Also displayed in \cref{fig:full-optical-cl} are the $E$-mode expectation band powers for a \ih{supercluster} region similar to the SuperCLASS field, $P_b^{\rm th}$, computed as described in \cref{sec:power_spectra-simulations}. For clarity, the expectation band powers are shown as a dashed line joining the band powers. The $B$-mode expectation band powers are zero and are not shown in the figure.

\subsection{Radio-optical and radio-radio power spectra of DR1 region}
\label{sec:power_spectra-radio_optical_dr1}
\begin{figure*}
\includegraphics[width=\textwidth]{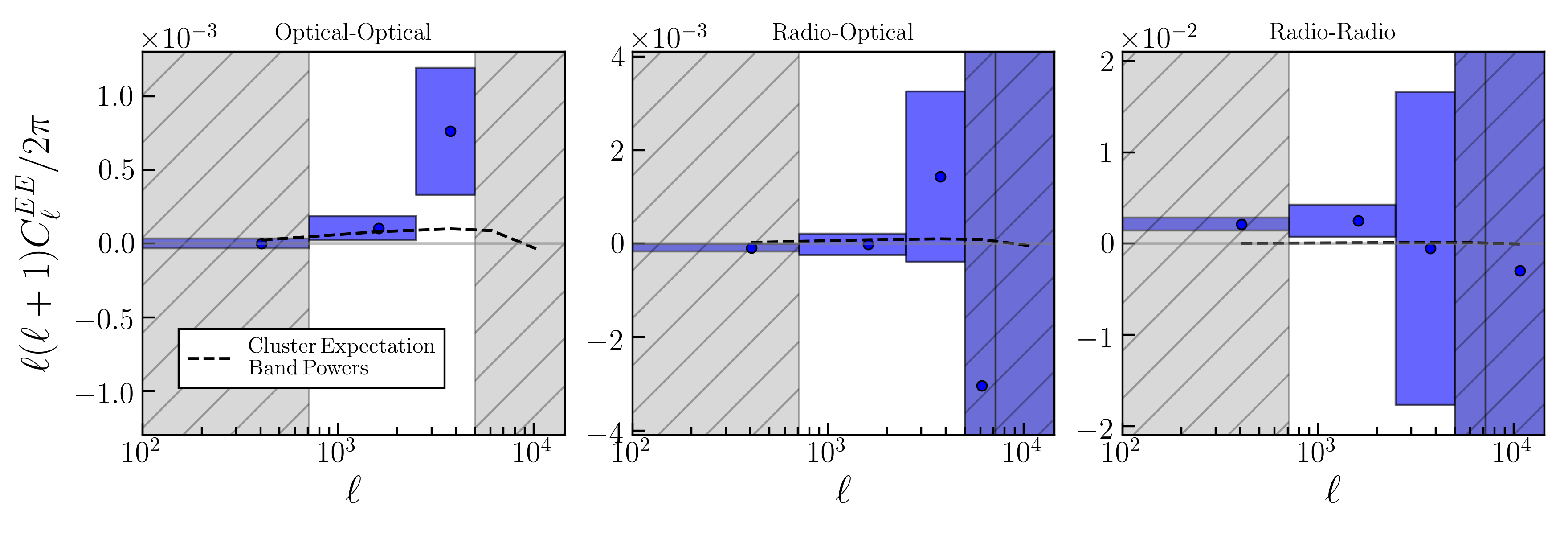}
\includegraphics[width=\textwidth]{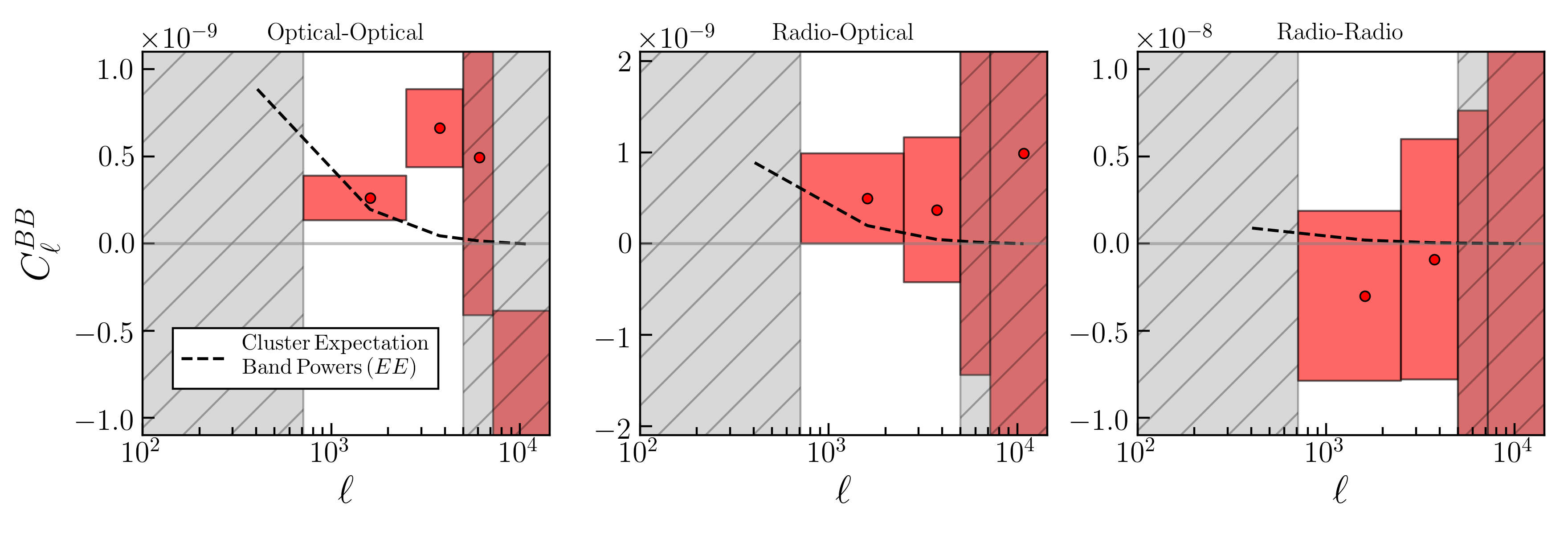}
\caption{Measured $E$- ({\it top row}) and $B$-mode ({\it bottom row}) power spectra for the $0.26\,\degsq$ DR1 region. From {\it left} to {\it right}, the panels display the optical-optical, radio-optical and radio-radio channels. Only the calibrated SuperCALS radio catalogue described in \cref{sec:radio_meas-shapes} was used to generate these power spectra, along with the same optical catalogue used for \cref{fig:full-optical-cl}. As in \cref{fig:full-optical-cl}, the $E$-mode cluster expectation band powers are illustrated by the dashed curve, but this time use the $0.26\, \degsq$ $\ell$-binning. Note the different vertical scaling between the $E$- and $B$-modes, as well as the different vertical ranges in each channel to cover the full extent of the error bars. Detection significances and signal-to-noise ratios are listed in Table~\ref{tab:power_spectra-DetSNR}.}
\label{fig:dr1-radio-optical-cl}
\end{figure*}

In \cref{fig:dr1-radio-optical-cl} we show the measured shear power spectra for the radio and optical catalogues in the $0.26\, \degsq$ DR1 region, including the optical-optical, radio-optical and radio-radio combinations. The top row shows the measured $E$-modes and the lower row the measured $B$-modes. As in \cref{fig:full-optical-cl}, the $E$-mode cluster expectation band powers are illustrated by the dashed curve, this time for the $0.26\, \degsq$ binning. Note the different vertical scales for each channel to contain the vertical extents of the error boxes. As one would expect, for this smaller $0.26\, \degsq$ region the random noise is too large to make a detection in any of the combinations, as is clear in \cref{fig:dr1-radio-optical-cl} and reported in Table~\ref{tab:power_spectra-DetSNR}.

\ih{We note that the band powers in \cref{fig:dr1-radio-optical-cl} were measured using a joint convergence of all three spectra simultaneously. We have also measured only the optical-optical spectra in the DR1 independently of the radio data, and find both methods to be consistent, and consistent with the optical-optical spectra in the full region in \cref{fig:full-optical-cl}.}

\begin{table*}
	\centering
	\caption{Detection significances, $D$ and signal-to-noise ratios, $S$ for the band power measurements shown in \cref{fig:full-optical-cl} and \cref{fig:dr1-radio-optical-cl}, calculated using \cref{eq:power_detection_significance} and \cref{eq:power_signal-to-noise}. Values calculated using the mean signal recovered from the simulated cluster data sets are also listed. All values were determined by only using the measurements of bands 2, 3 (and 4 for the full SuperCLASS region; see Table~\ref{tab:bandpowerselection}).
	For the DR1 $0.26\,\degsq$ rows, only the values measured using the calibrated SuperCALS radio shape catalogue are listed. For the DR2 $0.755\,\degsq$ Forecasts, the forecasts correspond to detections of a theory shear power spectrum given by the best-fitting band powers from the real optical data, see \cref{sec:power_spectra-DR2Forecasts}.
	}
	\label{tab:power_spectra-DetSNR}
	\begin{tabular}{c c c c c c c} 
		\hline
        Spectrum & Area & Figure & \multicolumn{2}{c}{Data} & \multicolumn{2}{c}{Simulations} \\
         & & & $D$ & $S$ & $D$ & $S$ \\
        \hline
        $\mathrm{OpOp}^{EE}$ & Full $1.53\,\degsq$ & \ref{fig:full-optical-cl} & \ih{9.9} & \ih{4.3} & \ih{6.4} & \ih{2.5} \\
        $\mathrm{OpOp}^{BB}$ & & & \multicolumn{2}{c}{\ih{4.6}} & \multicolumn{2}{c}{\ih{0.1}} \\
        \hline
        $\mathrm{OpOp}^{EE}$ & DR1 $0.26\,\degsq$ & \ref{fig:dr1-radio-optical-cl} & \ih{2.5} & \ih{2.2} & \ih{1.7} & \ih{1.1} \\
        $\mathrm{RadOp}^{EE}$ & & & \ih{0.8} & \ih{0.8} & \ih{0.5} & \ih{0.4} \\
        $\mathrm{RadRad}^{EE}$ & & & \ih{1.3} & \ih{1.4} & \ih{0.3} & \ih{0.3} \\
		$\mathrm{OpOp}^{BB}$ & & & \multicolumn{2}{c}{\ih{3.6}} & \multicolumn{2}{c}{\ih{0.2}}\\
		$\mathrm{RadOp}^{BB}$ & & & \multicolumn{2}{c}{\ih{1.1}} & \multicolumn{2}{c}{\ih{0.2}}\\
		$\mathrm{RadRad}^{BB}$ & & & \multicolumn{2}{c}{\ih{0.6}} & \multicolumn{2}{c}{\ih{0.2}}\\
		\hline
		$\mathrm{OpOp}^{EE}$ & DR2 $0.755\,\degsq$ & - & \multicolumn{2}{c}{-} & 6.0 & 3.3 \\
		$\mathrm{RadOp}^{EE}$ & Forecasts & & & & 2.2 & 1.6 \\
		$\mathrm{RadRad}^{EE}$ & & & & & 0.4 & 0.4 \\
		\hline
	\end{tabular}
\end{table*}

\subsection{DR2 Forecasts}
\label{sec:power_spectra-DR2Forecasts}

Table~\ref{tab:power_spectra-DetSNR} also presents forecasted detection significances and signal-to-noise ratios for simulated radio and optical weak lensing shape catalogues covering the DR2 area of $0.755\,\degsq$ -- the uniform depth area of the full VLA data set.

The forecasts assumed source number densities of $n^{\rm O}_{\rm{gal}} = 19$ gal arcmin$^{-2}$, as already measured and $n^{\rm R}_{\rm{gal}} = 1$ gal arcmin$^{-2}$, where the expected increase comes from the improved depth and resolution from the \emerlin + VLA combined image over the VLA-only one. For the optical sources, shape noise was sampled from the real data measurements, and for the radio sources, a Gaussian shape noise distribution was used with a standard deviation of $\sigma_\epsilon = 0.3$. Galaxy positions were randomly sampled within the $0.755\,\degsq$ area and the shear signals added to the shape catalogues, using the measured shear power from the full optical data (i.e. the data points in \cref{fig:full-optical-cl}). 61 pairs of simulations were generated and power spectra were extracted using a $0.755\,\degsq$ $\ell$-binning. As in \cref{sec:power_spectra-simulations}, the standard deviations of the output band powers were used as estimates of the error bars around the mean values for each $\ell$ bin. The forecasts show that the significance of the detection of the radio-optical power may be expected to be low, at $2.2 \sigma$.

\section{Demonstration of data combination from different telescopes}
\label{sec:data_combination}
As discussed in \cref{sec:radio_meas-shapes}, the combination of both VLA and \emerlin data will maximise the amount of available morphological information on the sources in the SuperCLASS survey. We take $uv$-plane approach to this data combination, which has the advantage over image-plane combination of the data \citep[e.g.\ ][]{2005MNRAS.358.1159M} that it delivers a well-defined, deterministic PSF for the {\sc clean}ed image, which is crucial for our analysis. Moreover, combination of data from both telescopes in the $uv$ plane may enable shape measurements to be made directly in the $uv$ plane, circumventing possible biases which may arise from the (non-linear) {\sc clean} process altogether.

\begin{figure*}
\includegraphics[width=0.975\textwidth]{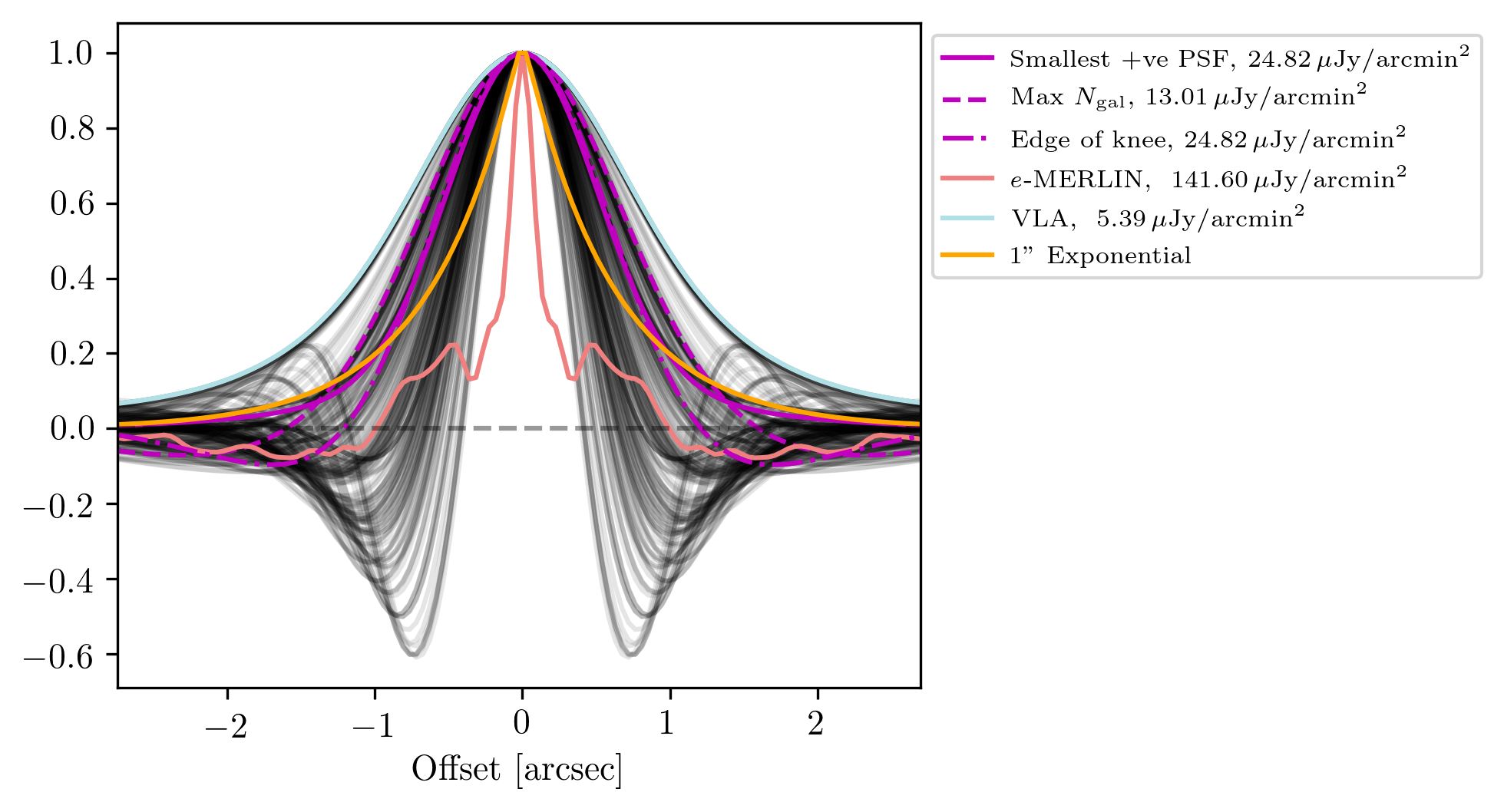}
\caption{Point spread functions for the various data weighting combinations considered. Individual telescope dirty beam PSFs are shown in blue (VLA) and red (\emerlin). Black lines show the dirty beam PSFs resulting from the different weightings tried, with those which were picked out for further exploration picked out in cyan. Numbers in the legend refer to the noise level in our simulated data set which was produced by each weighting. For reference we also show in orange a one arcsecond exponential \sersic profile.}
\label{fig:combined_psfs}
\end{figure*}

The SuperCLASS observations were designed with this data combination in mind, as the set of angular scales sampled by the VLA and \emerlin telescopes are highly complementary (see Paper I Fig. 1 and associated discussion). \emerlin has access to small angular scales ($\theta\sim 0.2''$) from widely separated antennae, but the lack of shorter separations means that much of the flux from diffuse sources falls on parts of the Fourier plane not covered by the telescope array and is `resolved out' (hence the problems with source detection discussed in \cref{sec:radio_meas-catalogue-emerlin}). The VLA antenna configuration provides a much denser sampling of small separations (large angular scales), and so more sensitivity to diffuse structure, but lacks sensitivity to small scales, with the smallest scale ($\theta\sim 1.5''$) sampled being larger than the $\sim1 \,$arcsec expected to be a typical size for sources relevant for shape measurement. These effects can be seen in \cref{fig:uvcoverage}, where the \emerlin coverage has large gaps at larger scales (small $u$ and $v$) and the VLA coverage contains no information at smaller scales (large $u,v$). \cref{fig:combined_psfs} also shows this information in real space in the form of the dirty beam PSFs for both VLA (blue, which is smooth and well behaved but broad) and \emerlin (red, which has a small central peak but complicated structure including negative sidelobes).

For the pointing labelled `J28' in our observation scheme (with pointing centre at 10:27:04.012, +68:09:27.00) we have performed a full combined imaging of our data from both VLA and \emerlin telescopes, with this data being at the full single image depth for both. In order to ensure scales in the image are always dominated by the array which contains the most signal on that scale, a weighting scheme is applied to the two data sets before combination, following a similar strategy to that used for the $e$MERGE Legacy Survey (Muxlow et al., 2020 in prep).

Rather than apply a global scaling of VLA or \emerlin data with respect to each other (equivalent to a step function in the $uv$ plane), our data combination approach allows us to apply a smooth tapering function throughout the $uv$ plane, smoothing out any shoulders in the combined-array PSF and enabling the trade-off between angular resolution and surface brightness sensitivity (which is inherent in \textit{every} interferometer dataset) to be explored. This weighting scheme consists of application of Tukey filters of differing widths to the visibilities; symmetric tapering functions which progressively downweight different parts of the outer and inner regions of the visibility plane. From an initial suite of 243 images made with different weighting functions (with beams shown by the grey lines in \cref{fig:combined_psfs}), we identify three for scientific exploitation. We use the recovered \clean beam major axis size and RMS noise for each image, and cut the T-RECS \citep{bonaldi-trecs} catalogue to find the total number of sources which will be resolved (have sizes larger than the beam major axis) and detected (have fluxes greater than five times the RMS noise value) by each choice of weighting. From this procedure we chose the `Max $N_{\rm gal}$' weighting as the one which returns the highest total of detectable, resolvable galaxies, and the `Edge of knee' weighting, which has a lower number of usable galaxies than Max $N_{\rm gal}$ but is smaller, sitting at the edge of a drop off in sensitivity with decreasing beam size. We also chose a third weighting according to the requirement that the dirty beam PSF is as small as possible without including negative sidelobes, which resolve out flux and can lead to problems in the imaging procedure. One dimensional cuts through the major axis of these beams are highlighted in purple in \cref{fig:combined_psfs}, along with a one arcsecond half-light radius \sersic exponential profile (shown in orange).

\begin{figure*}
\includegraphics[width=0.95\textwidth]{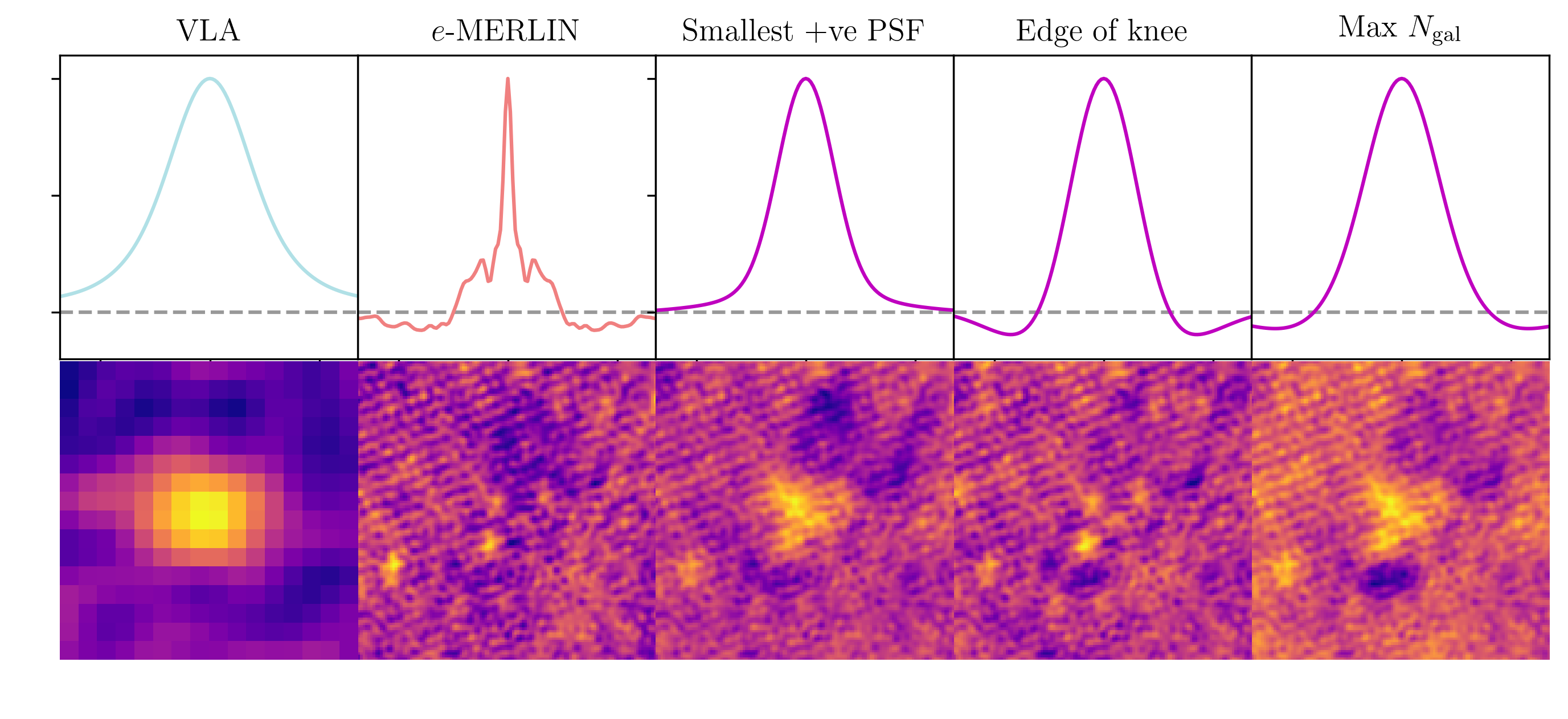}
\caption{\emerlin and VLA combined \clean images of a source in the J28 pointing, showing the differences in morphology information available from the different weightings, which are described in the text. For reference, one dimensional cut throughs of the dirty beam for each image are shown on the \emph{top} row.}
\label{fig:combined_image}
\end{figure*}

In \cref{fig:combined_image} we show the results of this data combination on deconvolved images of an example bright source, of simple morphology, chosen from our J28 data sets from \emerlin and the VLA. As can be seen, the unresolved emission in the VLA image appears undetectable in the \emerlin image, but is both detected and resolved in the combination images. We expect this effect will improve our shape measurements significantly for the full data release, by coherently including information on small angular scales from the \emerlin long baselines.


\section{Summary and conclusion}
\label{sec:summary}
We have analysed the first stage of data from the SuperCLASS experiment, consisting of $1.53\,\degsq$ of optical data from the Subaru telescope and $0.26\,\degsq$ radio data from the VLA and \emerlin telescopes. Previously existing radio surveys have not been designed with weak lensing in mind and so have not been capable of making a first detection of the signal. SuperCLASS has been designed as a survey of a region of sky containing a \ih{supercluster} at $z\sim0.2$ in the Northern sky with the aim of making the first detection of a weak lensing signal, both in the radio data alone and by cross-correlating the radio data with optical data. By making use of data from both the Karl G. Jansky VLA and \emerlin telescope arrays, SuperCLASS is sensitive to the range of Fourier scales expected to carry the morphological information about radio star-forming galaxies necessary to infer the shear signal from weak lensing. Here we have presented radio data from only the $0.26\,\degsq$ DR1 region, which contains information from the first $50\%$ ($\sim 400\,$ hours) of the \emerlin data, and is the area which is covered by these data to a uniform noise level of $7\,\mu$Jy/beam, along with data in this region from the full VLA data set.

Following \ih{Paper I} which describes the survey and the creation of the catalogues, we have further described the weak lensing methods applied to the data. For the Subaru Suprime-Cam optical data in the $BVr^\prime i^\prime$ bands we have used a pipeline consisting of well-proven methods and measured the two-point function of galaxy shapes, the non-zero signal in which we interpret as being due to the cosmic shear signal. As shown in \cref{fig:full-optical-cl} and \cref{tab:power_spectra-DetSNR} this allowed us to measure an $E$-mode shear power spectrum with a detection signficance of \ih{$9.9$} and signal-to-noise ratio of \ih{$4.3$}, confirming the presence of the lensing mass in the \ih{supercluster} region. We have additionally constructed other weak lensing observables from this data, including real space correlation functions (\cref{fig:corr_fns}), which are detected to be significantly larger than systematics signals estimated by cross-correlating the shapes of weak lensing sources (i.e. galaxies) with the shape of the deconvolved point spread function.

Using the radio catalogues described in \ih{Paper I} we have also measured the shapes of sources detected in deconvolved radio images. After performing a by-eye classification of sources, we identify those with low $\sim\mu$Jy fluxes and simple morphology as star-forming galaxies which may have their intensity profiles well-modeled by a \sersic profile with elliptical isophotes. In the deconvolved image from the VLA data we then use a method we call SuperCALS to measure the best fitting elliptical profile for each source. SuperCALS works by injecting simulated sources of known shapes into the residual image available from the \clean imaging process. The shapes measured for these injected sources are then used to form a model for the bias in shape in the true noise environment at that location in the image, and the real source measurements are corrected for this bias. We have constructed a sophisticated simulation pipeline, called SimuCLASS, consisting of sky model, simulated interferometer measurement and imaging reconstruction, which closely match the steps in the real data pipeline. Using this simulation pipeline, we have found that for the source population models (in terms of sizes, fluxes and profiles) expected for our observation, the VLA-only data does not provide the required resolution to successfully recover source shape measurements, but that the SuperCALS method does work adequately when the morphological information is available (by artificially increasing the size of the simulated sources). We have then applied the SuperCALS method to the real VLA data catalogue and formed the radio-radio and radio-optical ellipticity power spectra, finding no significant detection, as expected due to the low number density of sources leading to the signal being noise dominated -- we do not have enough galaxies to average down the `shape noise' from intrinsic galaxy ellipticities.

As discussed at length in Section 5 of \ih{Paper I} the \emerlin data alone `resolves out' much of the flux for the sources we wish to measure shapes with (much of the flux falls on parts of the Fourier plane not covered by the \emerlin telescope baselines). For the 56 sources which are detected in both the VLA and \emerlin images, we measure the \emerlin shapes with SuperCALS and provide them in the catalogue, but do not use them in our science analysis. Inclusion of \emerlin data will however allow improvements in the shape measurement when combined in a coherent way during the imaging process (i.e. through combination of data in the visibility plane). As discussed in \cref{sec:data_combination} we have begun this procedure, exploring different relative weighting schemes for the VLA and \emerlin data in the joint imaging process. The results of this section provide important information for the design of radio weak lensing surveys in the future: that their $uv$-coverage should be designed with extended source sensitivity, not point source sensitivity, in mind and with a dirty beam PSF which as closely as possible matches the expected source intensity profile.

By performing this data combination in the full data set consisting of all of the VLA data and a further $\sim 400\,$ hours of \emerlin data (which has been taken but not yet fully reduced) we will in the near future release results on a DR2 area covering a total area of $0.755\degsq$. If the data combination enables us, through reducing the effective noise in the image and improving the ellipticity measurement, to double the source density of radio sources to $1\,$arcmin$^{-2}$ then a marginal detection of a radio-optical cross power spectrum may be possible but is not expected, as shown in \cref{tab:power_spectra-DetSNR}, and a significant detection of a radio-radio power spectrum is unlikely.

In the future, weak lensing using radio data from the Square Kilometre Array (SKA) will be capable of cosmological constraints at the Stage III and Stage IV levels \citep[][]{2016MNRAS.463.3674H,2016MNRAS.463.3686B}, and will allow the formation of cross-correlations with optical surveys which will be highly robust to systematics \citep[][]{2017MNRAS.464.4747C}. This work represents a step forward in the sophistication of radio weak lensing and radio data analysis in general (e.g. through the use of the simulation pipeline and cross-correlation with optical data), and provides cutting-edge data on the flux distribution and morphology of star-forming galaxy radio sources at $\mu$Jy fluxes. The lessons learned, in particular on the importance of good telescope sensitivity across a wide range of angular scales, will be invaluable for future experiments such as the proposed $5\,\degsq$ VLA Deep Extragalactic Cosmology Survey (V-DECS) survey with the VLA and eventual surveys with the SKA.

\section*{Acknowledgments}
IH, MLB, SC and BT acknowledge the support of an ERC Starting Grant (grant no. 280127). TH is supported by a Science and Technology Facilities Council studentship. IH acknowledges support from the European Research Council in the form of a Consolidator Grant with number 681431, and from the Beecroft Trust. CAH acknowledges financial support from the European Union's Horizon 2020 research and innovation programme under the Marie Sk{\l}odowska-Curie grant agreement No 705332. We thank Joe Zuntz and Richard Rollins for useful discussions and assistance with \imshape, and Ian Smail for useful discussions.

CMC thanks the National Science Foundation for support through grants AST-1714528 and AST-1814034, and additionally CMC and SMM thank the University of Texas at Austin College of Natural Sciences for support. In addition, CMC acknowledges support from the Research Corporation for Science Advancement from a 2019 Cottrell Scholar Award sponsored by IF/THEN, an initiative of Lyda Hill Philanthropies. 
SMM also thanks the NSF for a Graduate Student Research Fellowship Program (GRFP) Award number DGE-1610403.

\bibliography{superclass2}
\bibliographystyle{mn2e_plus_arxiv}

\label{lastpage}

\end{document}